\documentclass[]{spie}  
\usepackage[]{graphicx}
\usepackage{hyperref}
\usepackage{wrapfig}
\usepackage[labelfont=bf]{caption}
\usepackage{floatrow}

\title{Photosensor Characterization for the Cherenkov Telescope Array: Silicon Photomultiplier versus Multi-Anode Photomultiplier Tube}

\author{A. Bouvier\supit{*,1}, L. Gebremedhin\supit{1}, C. Johnson\supit{1}, A. Kuznetsov\supit{1}, D. A. Williams\supit{1}\\
N. Otte\supit{2}, R. Strausbaugh\supit{2},
N. Hidaka\supit{3}, H. Tajima\supit{3},
J. Hinton\supit{4}, R. White\supit{4},
M. Errando\supit{5}, and R. Mukherjee\supit{5}
for the CTA consortium\supit{$\dagger$}
\skiplinehalf
{\small
\supit{1} Santa Cruz Institute for Particle Physics and Department of Physics, University of California, Santa Cruz, USA;\\
\supit{2}Center of Relativistic Astrophysics and School of Physics, Georgia Institute of Technology, USA;\\
\supit{3}Solar-Terrestrial Environment Laboratory, Nagoya University, Japan;\\
\supit{4}Department of Physics and Astronomy, University of Leicester, United Kingdom;\\
\supit{5}Department of Physics and Astronomy, Barnard College, Columbia University, USA\\
\supit{$\dagger$}\url{http://www.cta-observatory.org/?q=node/22}
\skiplinehalf
}
{\normalsize
\supit{*}send correspondence to: \href{mailto:aurelien.bouvier@gmail.com}{aurelien.bouvier@gmail.com}
}
}

\begin{document} 
  \maketitle

  \pagestyle{plain}
  \setcounter{page}{1}
  \pagenumbering{arabic}

\begin{abstract}

Photomultiplier tube technology has been the photodetector of choice for the technique of imaging atmospheric Cherenkov telescopes since its birth more than 50 years ago. Recently, new types of photosensors are being contemplated for the next generation Cherenkov Telescope Array. It is envisioned that the array will be partly composed of telescopes using a Schwarzschild-Couder two mirror design never built before which has significantly improved optics. The camera of this novel optical design has a small plate scale which enables the use of compact photosensors. We present an extensive and detailed study of the two most promising devices being considered for this telescope design: the silicon photomultiplier and the multi-anode photomultiplier tube. We evaluated their most critical performance characteristics for imaging $\gamma$-ray showers, and we present our results in a cohesive manner to clearly evaluate the advantages and disadvantages that both types of device have to offer in the context of GeV-TeV $\gamma$-ray astronomy.

\end{abstract}


\keywords{Silicon Photomultiplier, Photomultiplier tubes, Imaging Atmospheric Cherenkov Telescopes}

\section{INTRODUCTION}
\label{sec:intro}

\begin{figure}[t]
\centering
\includegraphics[height=3in]{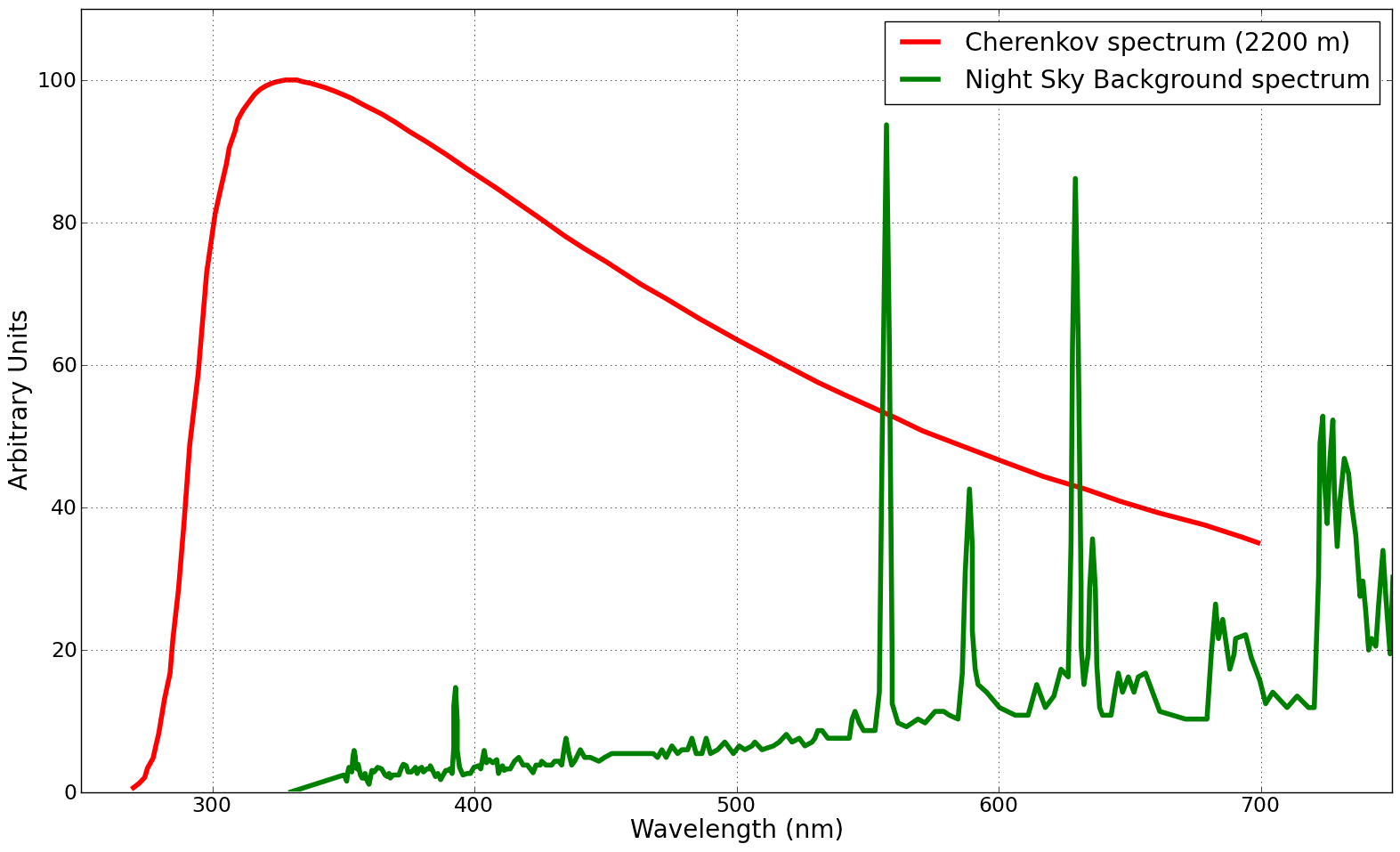}
\caption{Signal and background spectra for the IACT technique. {\bf Signal} (red curve): Observed Cherenkov spectrum from EAS at an altitude of 2200~m asl\cite{Doering01}. The Cherenkov emission spectrum scales as $\lambda^{-2}$ with atmospheric absorption cutting-off the spectrum at wavelengths below $\sim 300$~nm. {\bf Background} (green curve): Emission spectrum of the night sky background measured in La Palma\cite{Benn98}. Bright airglow emission lines above $\sim 550$~nm are mostly from atomic oxygen, hydroxide and sodium in the atmosphere. Note: units on the y-axis are arbitrary as the NSB normalization (and to a lesser degree its spectral features) strongly vary with site locations on Earth while the absolute normalization of the Cherenkov spectrum changes with the energy and impact distance of the EAS.}
\label{cherenkov_nsb}
\end{figure}

The imaging atmospheric Cherenkov telescope (IACT) technique aims at measuring the time, direction and energy of very high energy $\gamma$-rays (typically $\geq 100$~GeV) with the goal of studying the most energetic phenomena in the universe ({\it e.g.} blazars, pulsars, $\gamma$-ray bursts) as well as addressing fundamental questions about the laws of nature ({\it e.g.} the origin of dark matter, Lorentz invariance violation). This is accomplished by imaging flashes of Cherenkov light emanating from extensive air showers (EAS) that develop when a very high energy $\gamma$-ray or cosmic-ray interacts in the upper part of the Earth atmosphere. This Cherenkov emission has very distinctive temporal (lasting $\sim 5$~ns) and spectral (blue-ultraviolet peak) features that should be discriminated from the unavoidable night sky background (NSB) emission (see Figure \ref{cherenkov_nsb}). For this purpose, a fast and high-amplification photodetector sensitive in the $\sim 300-600$~nm range should be selected to populate the focal planes of telescopes with large light collection area. Photomultiplier tubes (PMTs) satisfy these requirements and have been the choice of predilection in this field since the early days.

Current generation IACT arrays (H.E.S.S.\cite{HESS04}, MAGIC\cite{MAGIC10} and VERITAS\cite{VERITAS08}) are all using a matrix of several hundreds ($\sim 500-2000$) of PMTs to populate cameras which are located at the focal plane of a Davies-Cotton (DC) telescope ({\it i.e.} with a multi-faceted single dish mirror). The Cherenkov Telescope Array\cite{CTA10} (CTA) will be an advanced observatory for ground-based $\gamma$-ray astronomy which is currently in its preparatory phase\cite{CTAintro13}. The plan calls for deployment of an array of 60 to 100 telescopes of different sizes and designs over $\geq 1$~km$^2$ in the southern hemisphere (the site location is still pending at this date). A second array of somewhat smaller scope is planned for in the northern hemisphere in order to cover the whole celestial sphere. As part of CTA, a two-mirror Schwarzschild-Couder (SC) telescope design\cite{Schwarzschild05} never built before is being envisioned for medium\cite{Vassiliev07,Vassiliev08} and small size telescopes. Advantages of this design are an improved (over DC design) optical point spread function over a wide field-of-view as well as isochronous optics allowing improved $\gamma$-ray angular resolution and rejection of cosmic-ray showers. One feature of the SC telescope design is its small plate scale which enables the use of compact photosensors and front-end electronics drastically decreasing the overall cost of the camera (thus compensating for the higher cost of the SC optics). This decreased plate scale opens the door to new categories of compact photosensors. The CTA consortium is currently considering two types of photosensors for its medium and small size Schwarzschild-Couder telescopes: the multi-anode photomultiplier tube (MAPMT) and the silicon photomultiplier (SiPM).

We performed in-depth characterization of both types of device in order to make an educated choice of the best photosensor to use for the Schwarzschild-Couder medium size telescope (SC-MST) for which the camera will consist of $\sim 11,000$ pixels, each with physical size $\sim 6 \times 6$~mm$^2$ (corresponding to $\sim 0.065 \times 0.065$~deg$^2$). Results of these investigations are presented in this paper highlighting the relevance of various device characteristics for the IACT application. We evaluated the performance characteristics that are most critical to this field: pulse shape (sec. \ref{sec:pulse_shape}), gain (sec. \ref{sec:gain}), temperature dependence (sec. \ref{sec:temperature}), light efficiency (sec. \ref{sec:efficiency}), noise (sec. \ref{sec:noise}) and aging (sec. \ref{sec:aging}). Although focus was put on the features that are most interesting to Cherenkov telescopes, we hope our in-depth study will prove useful to other applications which are considering the use of one of these photodetectors.

\subsection{Multi-Anode Photomultiplier Tube}
\label{sec:mapmt}

\begin{figure}[t]
\centering
\includegraphics[height=2in]{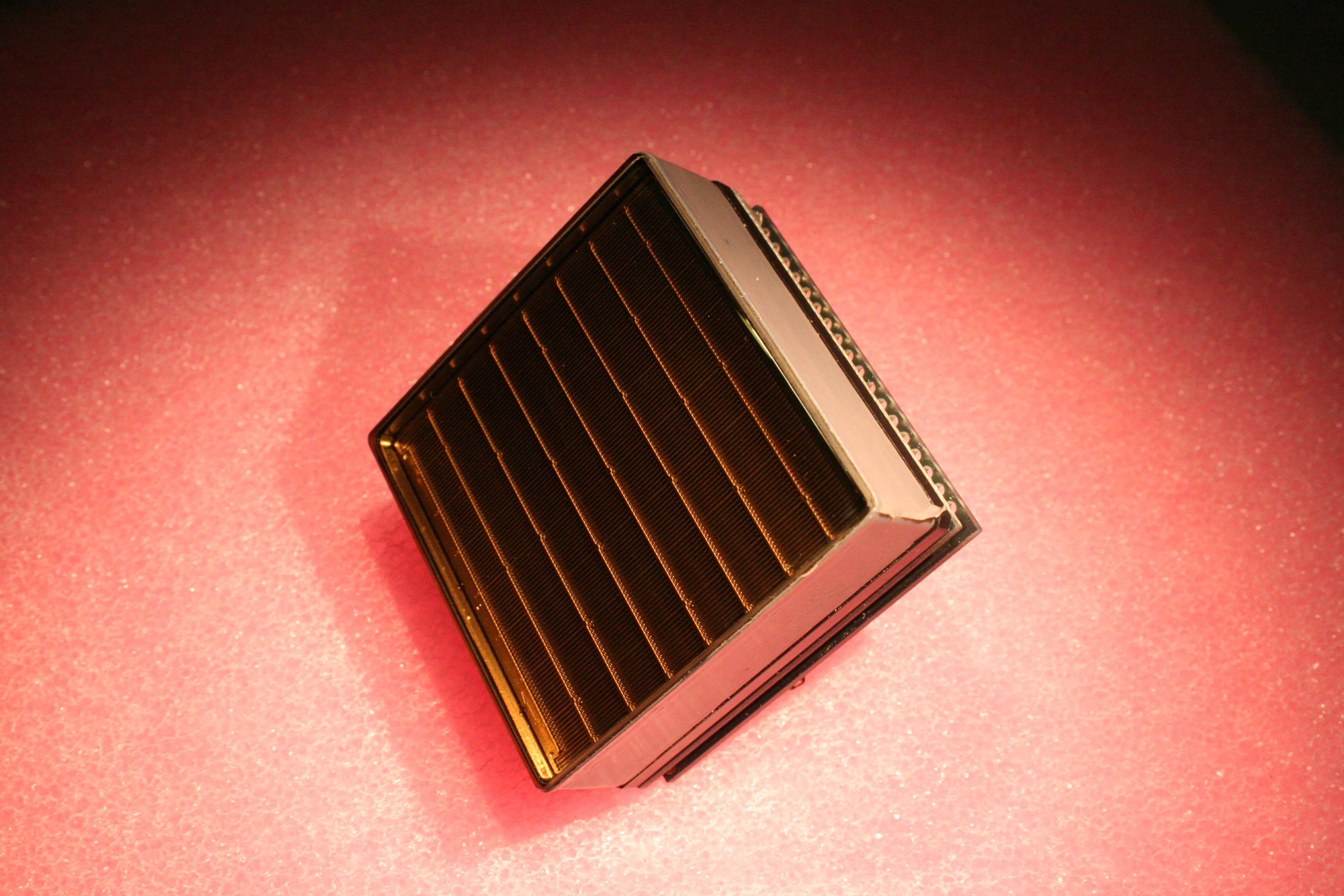}
\includegraphics[height=2in]{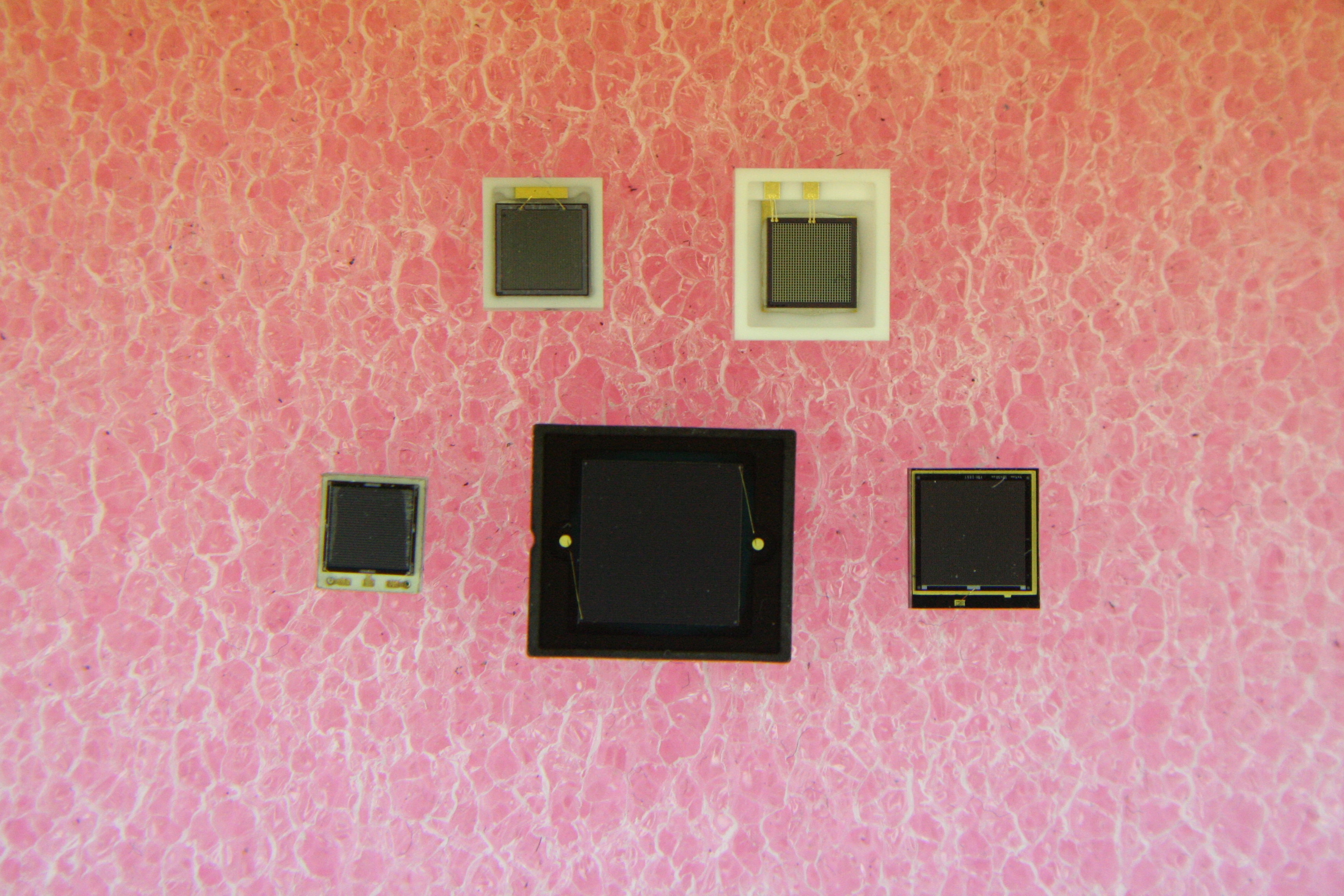}
\caption{{\bf Left}: Hamamatsu MAPMT H10966B: 8x8 pixel array. Module physical size is ~52x52~mm$^2$.  {\bf Right}: SiPM devices from five different manufacturers. From left to right, top to bottom: Excelitas C30742-33-050-C (series A), Hamamatsu 3x3-100C-LCT, Ketek PM3350, SensL P2MicroSB-30035-X13, FBK AdvanSiD NUV 4x4 mm$^2$. See table \ref{tab:sipm} for details.}
\label{mapmt_sipm_pics}
\end{figure}

The multi-anode photomultiplier tube is a position sensitive photodetector that is a direct extension of PMT technology. It is composed of a single photocathode layer underneath which resides a matrix of pixels each containing a dynode chain to amplify the initial photoelectron (PE) following the principles of PMT technology. In a sense, a MAPMT device is an array of small PMTs all contained within a single module and supplied by a single high voltage line. The compactness of the device is the attractive feature for our SC-MST design.
The MAPMT we considered in this work is the H10966B series from Hamamatsu Photonics\footnote{\url{http://www.hamamatsu.com/resources/pdf/etd/H8500_H10966_TPMH1327E02.pdf}} (Figure \ref{mapmt_sipm_pics}, left panel).

The H10966B is an 8x8 matrix with physical dimension of 52x52~mm$^2$ and 49x49 mm$^2$ of effective area. Pixel rows are 6.26~mm wide at the edge and 6.08~mm elsewhere. The photocathode of the device tested is made of super-bialkali material for higher photon detection efficiency and the entrance window is made of special UV glass which improves transmission in the ultraviolet part of the spectrum. Each pixel is composed of 8 dynodes for which voltage potentials are supplied by a single high voltage line.

Thanks to years of experience with PMTs, MAPMTs are a fairly familiar option and a natural continuation for the next generation Cherenkov telescopes. However, PMT technology comes with non-negligible drawbacks:

\begin{itemize}
\item Fragility (sealed vacuum tube)
\item Operation under high voltage (typically $\sim 1000-1500$~Volts)
\item Aging (due to heavy bombardment of the last dynode) 
\item Limited photon detection efficiency
\item Sensitivity to Earth magnetic fields
\item Large afterpulses
\end{itemize}

SiPMs have recently grown into a viable alternative technology which can address all of these issues.

\subsection{Silicon Photomultiplier}
\label{sec:sipm}

SiPMs\footnote{We note that Hamamatsu names its SiPM devices Multi-Pixel Photon Counter (MPPC) so we will use this nomenclature to refer to Hamamatsu devices.} are one of the most recent products in a long line of development in the field of fast semiconductor photosensors\cite{Dolgoshein03}. It is essentially a finely pixelated matrix of Geiger-mode avalanche photodiodes (GM-APDs) where each GM-APD cell is a reverse-biased PN junction which operates above breakdown voltage ($V_{breakdown}$). In this mode, photon or thermal excitation in the depleted region will produce a pair of charge carriers (electron-hole) which through impact ionization can trigger an electron-hole avalanche saturating the active area. Passive quenching is employed to stop the avalanche and reset the cell to a quiet state so that it becomes photosensitive again. Typical cell size in a SiPM ranges from $\sim$ 20 to 100~nm with thin quenching resistors deposited on top of the silicon substrate. Key advantages that SiPMs have been praised for are:

\begin{itemize}
\item Ruggedness
\item Low voltage operation (typically $\sim 20-100$~Volts) 
\item Low power consumption ($\leq 50 \mu$W/mm$^2$)
\item Resistance to high light levels (critical to continue observation during bright moonlight periods with Cherenkov telescopes)
\item High photon detection efficiency in principle achievable (although technically challenging, see sections \ref{sec:pde} and \ref{sec:noise})
\item Excellent pulse height resolution
\item Insensitivity to magnetic fields
\item Rapidly decreasing cost per mm$^2$
\end{itemize}

SiPMs stand as a very promising technology to replace current PMT-based cameras and there are ongoing efforts to study how well they could fulfill CTA requirements( see \cite{Mirzoyan13} for a recent review). The First G-APD Cherenkov Telescope (FACT) recently pioneered the use of SiPMs for the IACT technique by building a camera made of MPPC S10362-33-050C devices at the focal plane of a Davies-Cotton telescope\cite{FACT13}.

We have been collaborating with multiple SiPM manufacturers ({\it Excelitas}, {\it FBK AdvanSiD}, {\it Hamamatsu Photonics}, {\it Ketek} and {\it SensL}) to evaluate SiPM devices that would be most suitable for the IACT technique. The right panel of Figure \ref{mapmt_sipm_pics} shows some device samples from each of the manufacturers and table \ref{tab:sipm} provides details on all of the SiPM devices we are presenting results for in this paper.

\begin{table}[h]
\caption{List of SiPM devices which results are presented in this paper.} 
\label{tab:sipm}
\begin{center}       
\begin{tabular}{|c|c|c|c|c|c|c|} 
\hline
\rule[-1ex]{0pt}{3.5ex}  Manufacturer & Model number & Pixel size & Cell size & Coating & Trenches & Packaging \\
\hline
\rule[-1ex]{0pt}{3.5ex}  Hamamatsu & S10943-1071 & 3x3~mm$^2$ & 100~$\mu$m & Epoxy & No & 4x4 pixel array  \\
\hline
\rule[-1ex]{0pt}{3.5ex}  Hamamatsu & 3x3-100C-LCT & 3x3~mm$^2$ & 100~$\mu$m & None & Yes & Single chip  \\
\hline
\rule[-1ex]{0pt}{3.5ex}  Excelitas & C30742CERH-100-5-1 & 5x5~mm$^2$ & 100~$\mu$m & Epoxy & No & Single chip  \\
\hline
\rule[-1ex]{0pt}{3.5ex}  Excelitas & C30742-33-050-C & 3x3~mm$^2$ & 50~$\mu$m & Epoxy & Yes & Single chip  \\
\hline
\rule[-1ex]{0pt}{3.5ex}  SensL & P2MicroSB-30035-X13 & 3x3~mm$^2$ & 35~$\mu$m & Epoxy & No & Single chip  \\
\hline
\rule[-1ex]{0pt}{3.5ex}  Ketek & PM3350 & 3x3~mm$^2$ & 50~$\mu$m & Epoxy & Yes & Single chip  \\
\hline
\end{tabular}
\end{center}
\end{table} 

\section{Single Photon Response}
\label{sec:single_pe_response}

MAPMT and SiPM are both high amplification devices which are capable of detecting single photons. We describe in this section the typical single photon response from these devices in terms of pulse shape and internal gain.

\subsection{Pulse Shape}
\label{sec:pulse_shape}

\begin{figure}[t]
\centering
\includegraphics[height=2.4in]{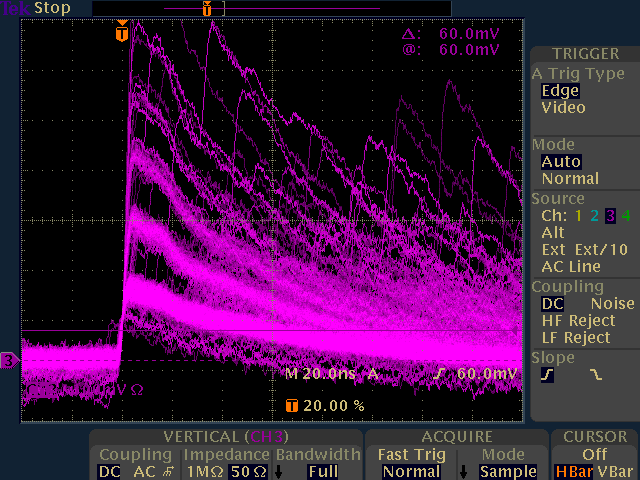}
\includegraphics[height=2.5in]{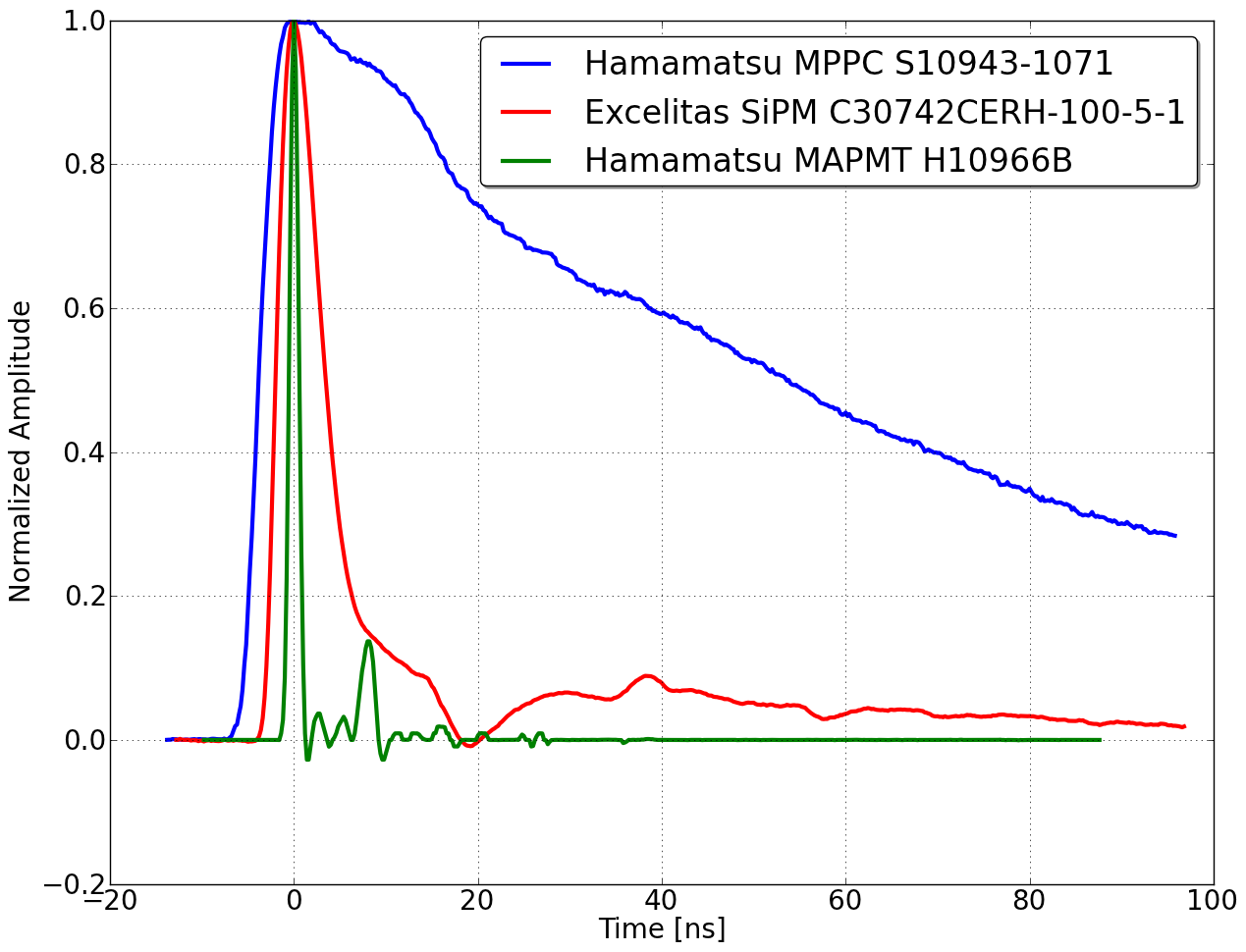}
\caption{{\bf Left}: Output signal from Hamamatsu MPPC S10943-1071. The trigger threshold is set at $\sim 0.5$~PE level. The clear distinction between the separate PE levels (1, 2, 3, 4, 5~PEs) illustrates the excellent pulse height resolution of SiPM devices (see section \ref{sec:gain}). SiPM cross-talk is the reason behind  pulses with amplitudes of 2~PEs and above (see section \ref{sec:xtalk_sipm}). {\bf Right}: Average pulse shape from MPPC S10943-1071 ($FWHM \sim 60$~ns, $t_{rise} \sim 6$~ns), Excelitas C30742CERH-100-5-1 ($FWHM \sim 6$~ns, $t_{rise} \sim 3$~ns) and MAPMT H10966B ($FWHM \sim 1.5$~ns, $t_{rise} \sim 1$~ns)}
\label{pulse_shape}
\end{figure}

A MAPMT or SiPM avalanche triggered by a single photon (or thermal excitation) will produce a fast, high amplitude voltage pulse. Figure \ref{pulse_shape} (left panel) for example shows typical output signals from MPPC S10943-1071. The average pulse shape for this device is shown in Figure \ref{pulse_shape} (right panel) and is compared to two other devices: Excelitas C30742CERH-100-5-1 and MAPMT H10966B. The MAPMT is the fastest device of them all with a full width at half maximum (FWHM) of $\sim 1.5$~ns due primarily to the time spread accumulated by avalanche electrons during their transit from the photocathode to the anode. In comparison, the SiPM pulse shape is significantly broader with FHWM ranging from $\sim 5$ to $\sim 100$~ns depending on the manufacturer. The rise time ($t_{rise}$) however, remains short enough ($\leq 2-10$~ns depending on the device) that shaping circuitry acting as a high pass filter can be used to significantly reduce the FWHM to $\sim 10-20$~ns (although at the cost of attenuating the signal amplitude).

The Cherenkov signal from EAS that we are aiming to detect is typically $\sim 5$~ns long. Pulse widths much smaller than this intrinsic duration from the Cherenkov signal might therefore result in non-overlapping Cherenkov photon pulses which would fail to pass our trigger threshold (which we anticipate to be at the level of $\sim 2-4$~PEs). For example, MAPMT pulses may need to be artificially broadened to resolve this issue. On the other hand, the pulse width should not be much larger than the intrinsic Cherenkov pulse duration in order to limit the contamination from NSB photons during the Cherenkov signal pulse. Shaping of the signal output is therefore anticipated to shorten SiPM pulses with long tails.

\subsection{Gain}
\label{sec:gain}

\begin{figure}[t]
\centering
\includegraphics[height=2in]{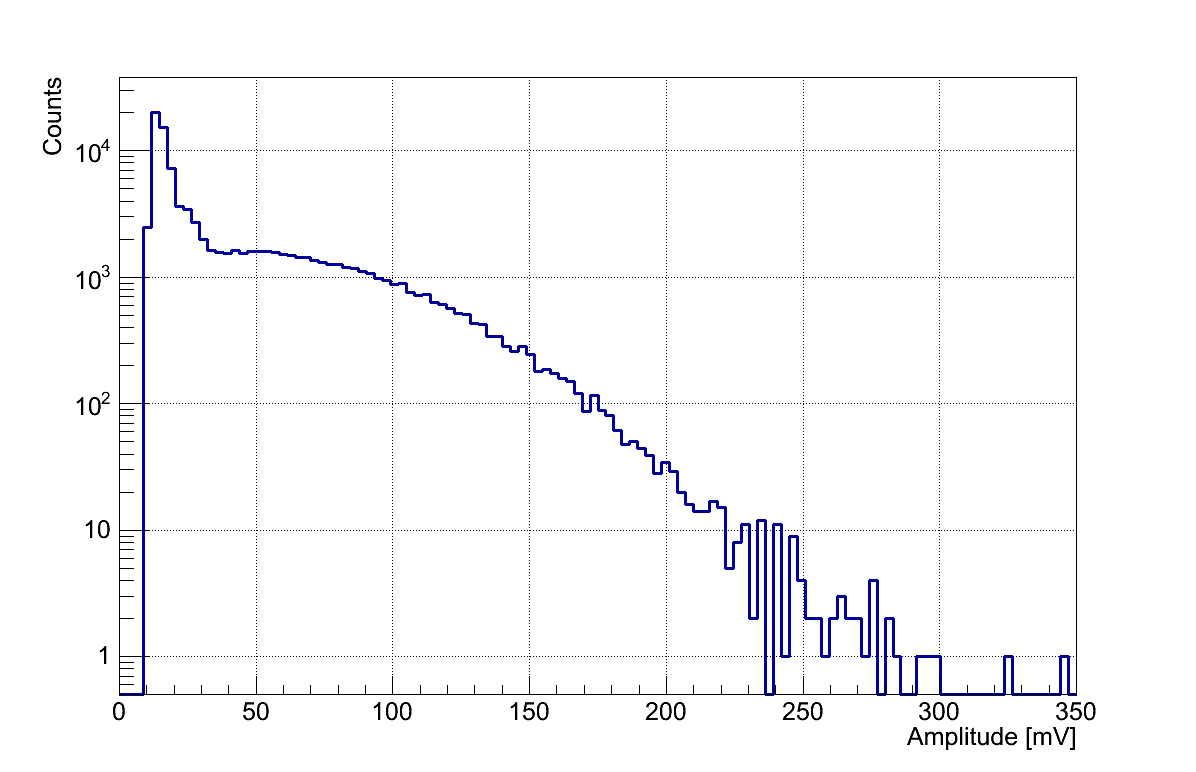}
\includegraphics[height=2in]{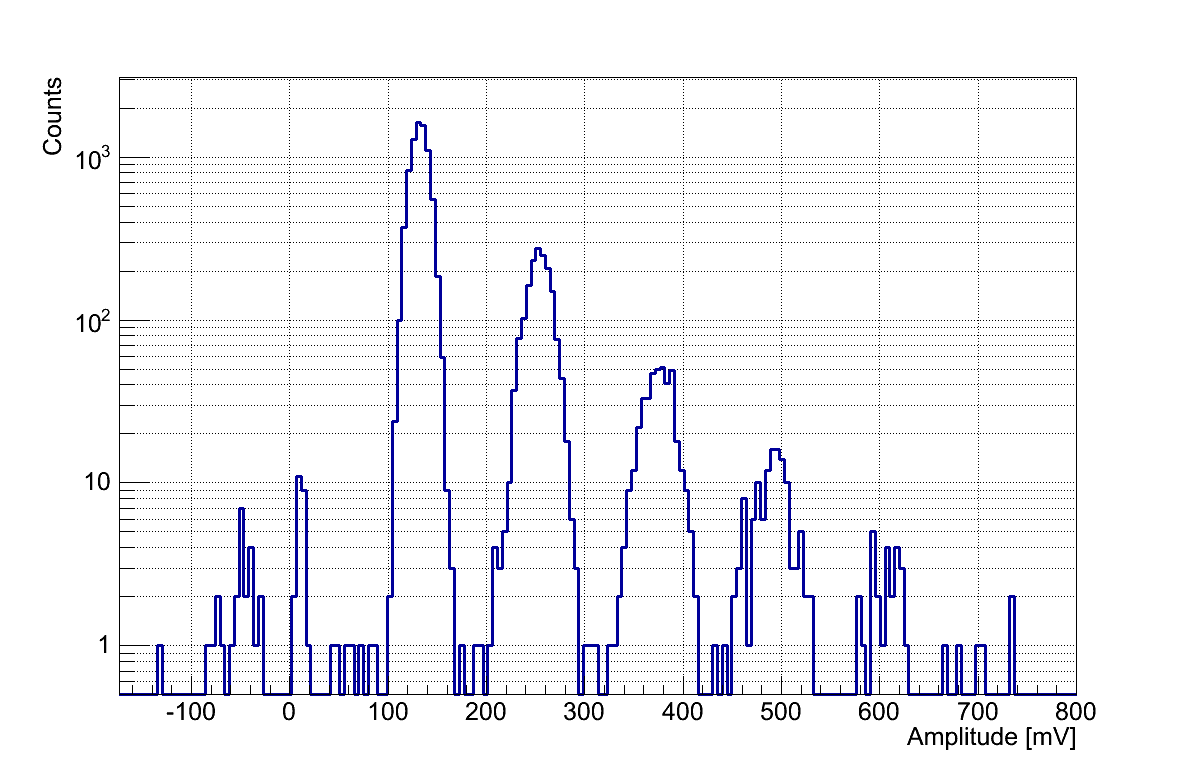}
\caption{Pulse height distribution of MAPMT H10966B at 890~volts  (left) and MPPC S10943-1071 at 5$^{\circ}$C and 71.3~volts (right). Because MPPC data acquisition system was triggered on the signal output ($\sim 0.5$~PE threshold), no pedestal peak appears in its pulse height distribution. MAPMT H10966B however was triggered coincidently with dim LED flashes ($\leq 0.05$~PE per pulse) so its pulse height distribution includes the pedestal and 1~PE peak ($\leq 5\%$ level contamination from 2~PE peak). The right panel illustrates the excellent single photoelectron resolution of SiPM devices with clear resolution of the 1, 2, 3, 4 and even 5~PE peaks. We note that the peak amplitudes include the electronic gain amplification which in this case was $\sim 144$ (MAPMT H10966B) and $\sim 524$ (MPPC S10943-107). }
\label{pulse_height}
\end{figure}

Gain is the internal amplification of the device expressed as the average number of charge carriers produced from a single original photoelectron (or thermal electron). It can be estimated from the mean voltage amplitude of the 1~PE peak $A_{av,peak}$ (see Figure \ref{pulse_height}) with:

\begin{equation}
\label{eq:gain}
Gain = \frac{A_{av,peak}}{50\textrm{ ohm}} \times \frac{C_{current=>charge}}{G_{elec} \times e}
\end{equation}

\noindent
where $C_{current=>charge}$ is the conversion factor from peak current to total integrated charge (using the average pulse shape), $G_{elec}$ is the amplification of the external electronic gain chain used and e is the electron charge.

Gain strongly correlates with MAPMT high voltage or SiPM bias voltage $V_{bias}$ (Figure \ref{gain}, top left panel). The linear dependence of SiPM gain with $V_{bias}$ can be simply understood by modeling each GM-APD depleted region with a capacitor of capacitance $C_{cell}$. The charge flowing during capacitor discharge is then:

\begin{equation}
\label{eq:capacitance_charge}
Q = Gain \times e = C_{cell} (V_{bias} - V_{breakdown})
\end{equation}

We note that the slope of the Gain versus $V_{bias}$ curve can be used to derive the equivalent cell capacitance $C_{cell}$. For Excelitas C30742CERH-100-5-1 (Figure \ref{gain}), for example, we find $C_{cell} \sim 350$~fF which is fairly typical of SiPM cell capacitance.

Thanks to the excellent cell-to-cell uniformity achieved during manufacturing of a SiPM pixel, the pulse-to-pulse SiPM gain fluctuations (quantified by the width of the single PE peak) are extremely small. This is for example illustrated by the clear resolution of the different photoelectron peaks in the right panel of Figure \ref{pulse_height}.
On the other hand, the MAPMT has large pulse-to-pulse gain fluctuations which result in a broad single PE peak which is difficult to resolve from the pedestal peak (Figure \ref{pulse_height}, left panel). This is mainly caused by Poisson fluctuations in generating secondary electrons at the first dynode stage (typical single dynode amplification being a factor of $\sim 4-6$). Increasing the high voltage increases the amplification at the first dynode and improves the single PE peak resolution, but only slightly. Since operating at lower gain is usually desirable to limit aging (see Section \ref{sec:aging}), this is generally not an option.

This lack of resolution of the MAPMT single PE peak causes large uncertainty in the gain calibration of individual pixels (because of large uncertainty on the factor $A_{av,peak}$ in Equation \ref{eq:gain}). This directly impacts the energy resolution of Cherenkov telescopes as the photosensor gain needs to be calibrated out to estimate the Cherenkov pulse intensity and $\gamma$-ray energy. For high energy and small impact distance $\gamma$-rays the number of Cherenkov photons is large enough that gain fluctuations will average out to leave only a small uncertainty on the $\gamma$-ray energy estimate. However, only few Cherenkov photons will be detected for faint or distant $\gamma$-ray showers, which therefore suffer from this effect, increasing the energy uncertainty close to the instrument energy threshold.

\section{Temperature dependence}
\label{sec:temperature}

\begin{figure}[t]
\centering
\includegraphics[height=2in]{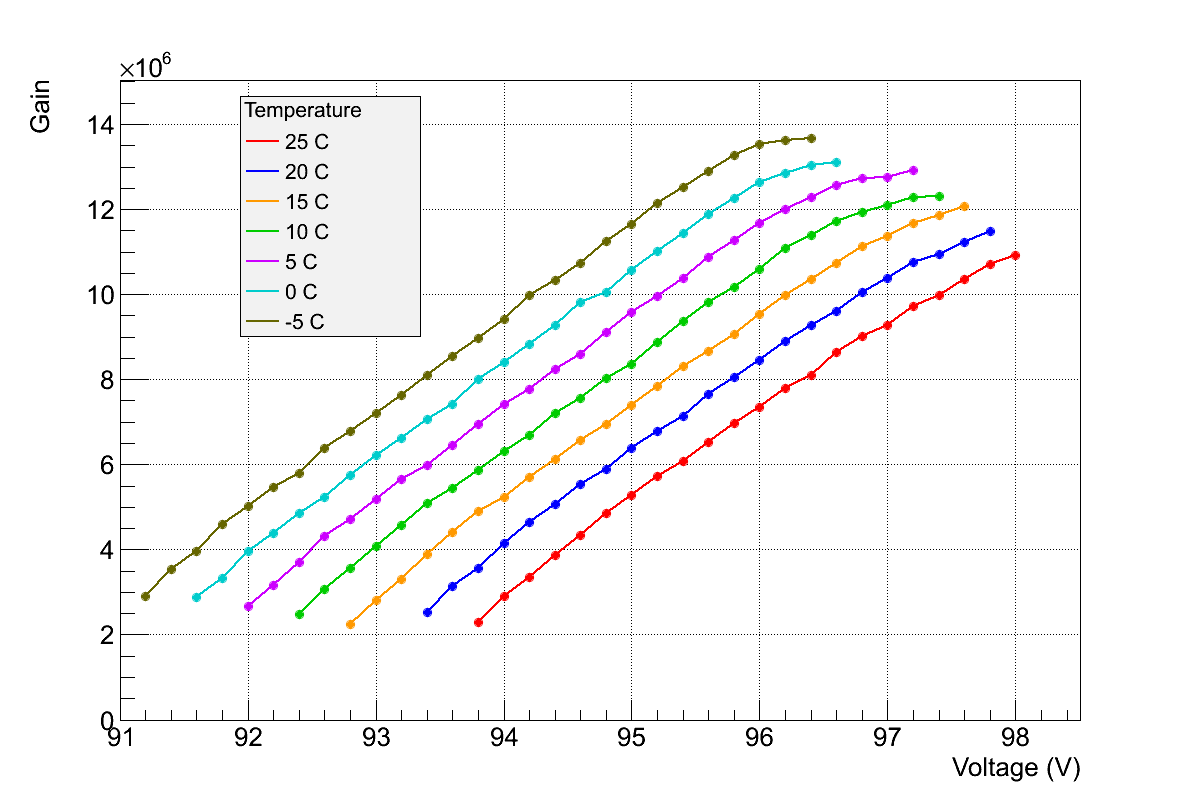}
\includegraphics[height=2.1in]{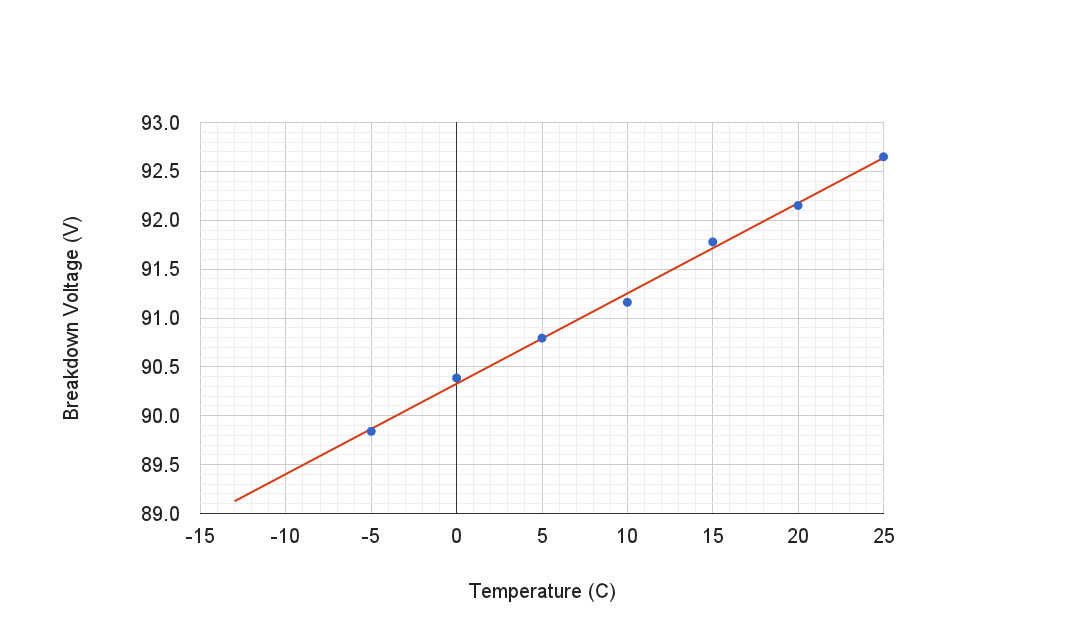}
\includegraphics[height=2.3in]{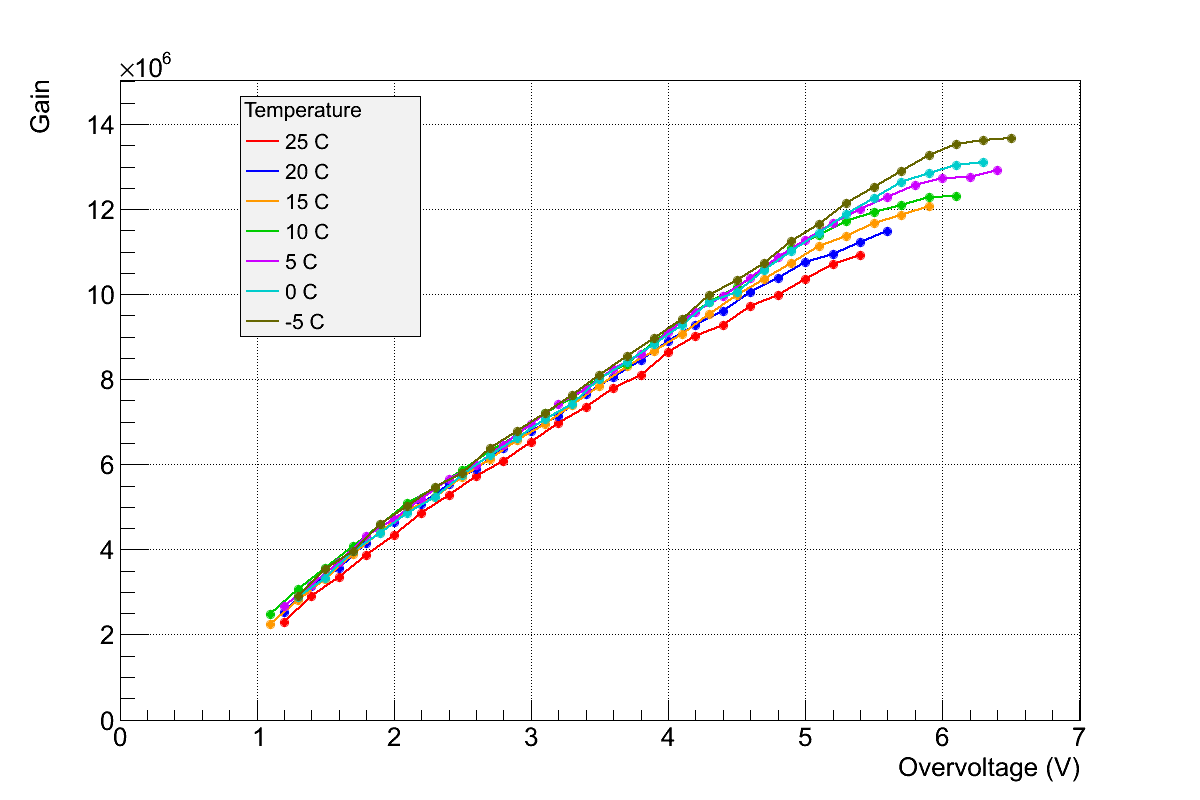}
\caption{Gain measurements on Excelitas C30742CERH-100-5-1 at different temperatures (-5~$^{\circ}$C to 25~$^{\circ}$C) and bias voltages. {\bf Top left}: Gain as a function of $V_{bias}$ shows strong temperature dependence due to $V_{breakdown}$ intrinsic temperature dependence. $V_{breakdown}$ can be estimated from a linear extrapolation of each curve toward zero gain. {\bf Top right}: $V_{breakdown}$ as a function of temperature for Excelitas C30742CERH-100-5-1. Linear fit yields a slope of $\sim 93$ mV/$^{\circ}$C for this device. {\bf Bottom}: Gain is essentially temperature independent when parametrized with overvoltage. The gain roll-over at high overvoltages is the result of voltage drop across the bias resistor which occurs when high currents flow through the SiPM. This effect occurs at lower overvoltages for higher temperatures because of higher dark rates.}
\label{gain}
\end{figure}

A major source of worry with SiPM devices is their high temperature sensitivity. The top left panel of Figure \ref{gain} shows the gain of Excelitas C30742CERH-100-5-1 as a function of $V_{bias}$ and temperature and a clear temperature dependence can be observed. Looking at Equation \ref{eq:capacitance_charge}, it appears $V_{breakdown}$ can be inferred from the intersection of the linear extrapolation of each curve with $Gain = 0$. Using this method, Figure \ref{gain} (top right panel) shows $V_{breakdown}$ linear dependence with temperature for this chip ($\sim 93$ mV/$^{\circ}$C). The typical range for this temperature dependence on SiPM devices we tested is $\sim 50-100$~mV/$^{\circ}$C.

One can correct for this intrinsic temperature dependence of $V_{breakdown}$ by parametrizing SiPM performance as a function of overvoltage ({\it i.e.} $V_{bias} - V_{breakdown}$).
The bottom panel of Figure \ref{gain} shows that the gain of Excelitas C30742CERH-100-5-1 indeed becomes temperature independent when parametrized with overvoltage. We find this to be true for the large majority of SiPM devices. The main parameter that still remains temperature dependent even as a function of overvoltage is dark noise (see Section \ref{sec:dark_noise}).

In practice, it is critical to operate SiPM devices at constant overvoltage in order to keep the device performance (gain, efficiency, cross-talk) unchanged throughout instrument operation. We envision two ways of implementing this in the SC-MST telescope: bias voltage regulation (based on calibration of $V_{breakdown}(T)$) and temperature control.

\section{Light Efficiency}
\label{sec:efficiency}

High efficiency at detecting photons is the primary goal of a Cherenkov telescope in order to reveal faint $\gamma$-ray showers and thus achieve the lowest energy threshold possible as well as extend the sensitivity to showers with high impact distances. Several strategies can be undertaken to maximize efficiency at detecting Cherenkov light: use of mirrors with large collecting area and high reflectivity, minimize shadowing from the optical support structure, fill the focal plane with high efficiency photosensors that use a layout which minimizes dead areas. In this section, we will look at the later two items that are related to the photosensor component.

\subsection{Photon Detection Efficiency}
\label{sec:pde}

The photon detection efficiency (PDE) is the probability for an incoming photon to produce a detectable electronic signal. There are multiple processes that need to happen for it to occur and we will now describe each of them in the case of MAPMT (sec. \ref{sec:pde_mapmt}) and SiPM (sec. \ref{sec:pde_sipm}).

\subsubsection{Photon detection with MAPMT}
\label{sec:pde_mapmt}

Below are the processes involved in the detection of light with a MAPMT device:

\begin{enumerate}

\item A photon first has to transmit through the air-glass-photocathode optical interface. Reflection alone generally accounts for $\sim 10-30\%$ loss although anti-reflective layers can reduce this number. Glass absorption is the highest in the UV and H10966B uses a special UV glass which effectively reduces ultraviolet absorption during photon propagation through the glass.

\item Once inside the super-bialkali material, the photon needs to excite an electron from the valence to the conduction band (a wavelength dependent process). The free electron will then random walk within the semiconductor conduction band and has a certain probability to reach the photocathode-vacuum interface with enough kinetic energy to overcome the work function and escape into the vacuum. The combined probability of these first two steps is generally referred to as quantum efficiency (QE).

\item The photoelectron in the vacuum can then be accelerated by electric fields, focused onto the first dynode and emit secondary electrons. The probability for this whole step to happen is usually referred to as the collection efficiency. Three main processes contribute to the losses here: the photoelectron misses the first dynode, the photoelectron is elastically backscattered and the photoelectron hits the first dynode without producing any secondary electrons.
The stronger the amplitude of the electric field, the higher the collection efficiency will be. It is therefore a desirable feature to set the voltage between the photocathode and the first dynode as high as practically feasible (an upper limit is generally imposed to prevent arcing inside the tube). In the case of H10966B, such optimum voltage is achieved for a total high voltage of 1200~V which results in $\sim 115$~V between the photocathode and first dynode\cite{Hamamatsu_private}.

\item Secondary emission electrons will then be produced at each dynode stage and further accelerated toward the next stage by the electric field lines to finally produce a high-amplification current pulse onto the anode. This step does not suffer from any significant efficiency loss as the number of secondary electrons is high enough even at the first dynode that the chance for all of them not to be collected onto the second stage is quite slim.

\end{enumerate}

We note that step 2 in this list requires a compromise in the photocathode thickness because a thick layer increases the chance of photon absorption while a thin layer enhances the chance for the photoelectron to escape into the vacuum. This compromise intrinsically limits the highest PDE values that can be achieved with the PMT technology. In principle, this limitation could be overcome by amplifying the signal directly inside the semiconductor material where the photoelectron is produced (the photoelectron would not need to escape the semiconductor in such case). This is exactly what GM-APDs and therefore SiPMs achieve.

\subsubsection{Photon detection with SiPM}
\label{sec:pde_sipm}

SiPM PDE can be summarized by the following formula:

\begin{equation}
PDE =  P_{transmission} \times QE \times P_{avalanche} \times \epsilon_{fill\textrm{ }factor}
\end{equation}

\noindent
where:

\begin{enumerate}

\item $P_{transmission}$: transmission probability for a photon to enter the depleted region. Bare silicon has a high surface reflectivity ($\sim 30 \%$). To improve transmission probability, it is usual practice to deposit on the silicon substrate a well chosen thickness of coating with refraction index intermediate between that of air ($n_{air} \sim 1$) and the silicon substrate ($n_{Si} \sim 3.4$) as well as high transparency. Epoxy ($n_{epoxy} \sim 1.55$) or silicone gel ($n_{silicone\textrm{ }gel} \sim 1.4$) are commonly used for example. Silicone gel having a higher UV transparency, it is therefore a preferred choice for our application. The coating also provides protection and insulation to the micro-cells, aluminum traces and wire bonding part of the SiPM pixel. We note that additional anti-reflective layers are sometimes used by manufacturers.

\item $QE$: probability for a photon to excite an electron from the valence to the conduction band inside the depleted region, thus creating a pair of charge carriers (electron-hole). This step is responsible for most of the wavelength dependence of the PDE and is referred to as quantum efficiency.

\item $P_{avalanche}$: probability for an electron-hole pair to initiate an avalanche and saturate the active area inside the micro-cell (so-called Geiger mode discharge). This process strongly depends on the electric field inside the depleted region which is why the PDE correlates strongly with the applied overvoltage (Figure \ref{pde_sipm_mapmt}, left panel).

\item $\epsilon_{fill\textrm{ }factor}$: fraction of the SiPM pixel which is occupied by active areas of GM-APD cells. Dead areas are mostly the result of the poly-silicon quenching resistors and aluminum strips carrying out the signal from individual cells. This so-called fill factor can be maximized by using large cell size at the cost of lowering the pixel dynamic range and increasing the cell capacitance. We are currently in the process of determining whether 50~$\mu$m ($\epsilon_{fill\textrm{ }factor} \sim 60\%$) or 100~$\mu$m ($\epsilon_{fill\textrm{ }factor} \sim 80\%$) cell size would be better for our application. Finally, Hamamatsu is also developing a semi-transparent metal resistor technology with high transparency ($\sim 80\%$) that can be deposited somewhere on top of the cell resulting in an increased total fill factor.

\end{enumerate}

\subsubsection{Experimental measurement technique}

\begin{figure}[t]
\centering
\includegraphics[height=4in]{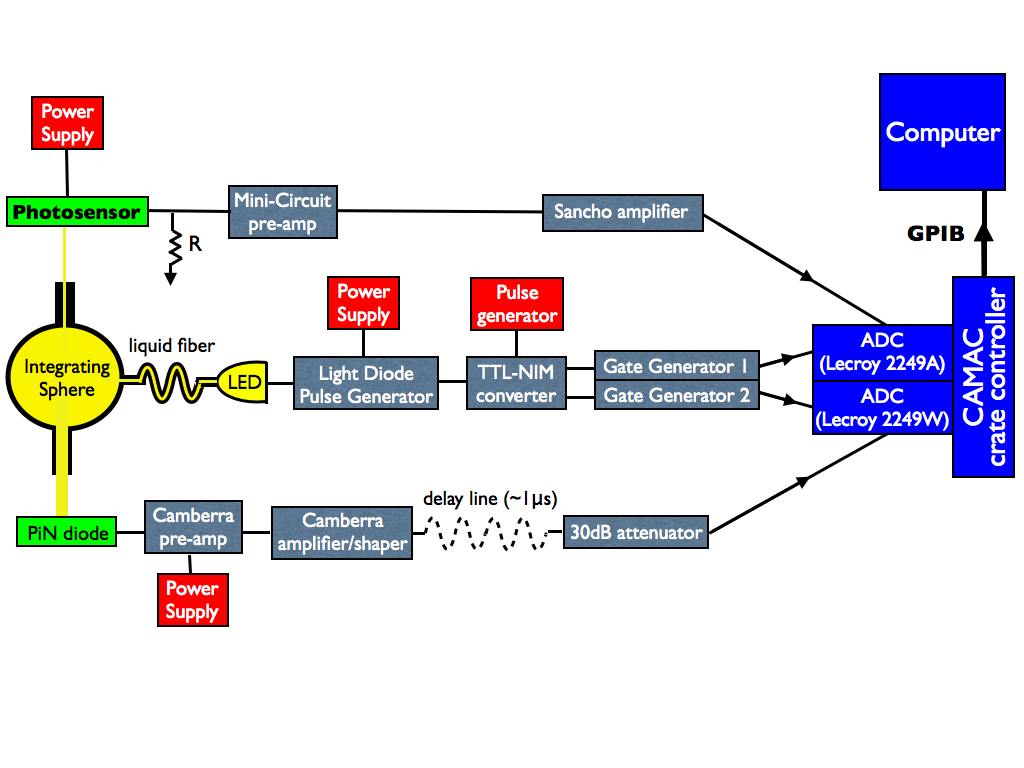}
\caption{Schematic of our PDE experimental setup. An LED is flashed into an integrating sphere which splits the light toward a calibrated PIN photodiode and the photosensor being tested.  Both signals are then amplified and integrated by analog to digital converter modules. A CAMAC crate is then routing the digital values to a computer which will perform the analysis described in the text.}
\label{pde_setup_schematic}
\end{figure}

Figure \ref{pde_setup_schematic} shows a schematic of our experimental setup which we use to measure the PDE of various photodetectors. The general method used is now fairly standard and has been described in several places (see \cite{Bondarenko98} or \cite{Otte06} for example). A flashing light source (an LED with narrow band filter in our case) is illuminating both a calibrated reverse-biased PIN photodiode (Hamamatsu S3590-18) along with the photodetector being tested. In our setup, we use an integrating sphere to accomplish this step. Calibration of the PIN diode quantum efficiency $QE_{PIN}$ (performed by Hamamatsu), digital count to photoelectron conversion $C_{DC=>PE}$ (by detecting the 59.5~keV $\gamma$-ray from $^{241}$Am), and splitting ratio SR of the two integrating sphere ports (by steady illumination of the sphere and measuring the light exiting both ports with a photodiode operating in current mode) allow us to determine the average number of photons hitting the photosensor during an LED pulse: 

\begin{equation}
<N_{ph,incoming}> = \frac{<DC_{PIN}> \times C_{DC=>PE}}{QE_{PIN}} \times SR
\end{equation}

\noindent
where $<DC_{PIN}>$ is the mean digital counts detected onto the PIN diode during an LED pulse. The signal from the photosensor is amplified and integrated over the duration of the LED pulse. A pulse height distribution is then constructed to derive the average number of LED photons detected per pulse: $<N_{ph,detected}>$. To do this, we first set an analysis threshold at $\sim 0.5$~PE level and define the following parameters:

\begin{itemize}
\item $N_{tot,ped}$: total number of events recorded during a dark run ({\it i.e.} no flashing LED)
\item $N_{0,ped}$: number of events below threshold during a dark run (this is not equal to $N_{tot,ped}$ mostly because of dark noise, see section \ref{sec:dark_noise})
\item $N_{tot,LED}$: total number of events recorded during an LED run ({\it i.e.} LED is flashing)
\item $N_{0,LED}$: number of events below threshold during an LED run
\item $N_{0,true}$: true number of events with no LED photons detected during LED run (can be derived from the parameters above)
\end{itemize}

The probability for a pulse without any LED photons detected to still have a signal above threshold (mainly the result of dark pulses) can be derived from the dark run as:

\begin{equation}
\label{eq:pthres}
P_{>threshold} = \frac{N_{tot,ped} - N_{0,ped}}{N_{tot,ped}}
\end{equation}

\noindent
which implies that:

\begin{equation}
\label{eq:n0true}
N_{0,true} = N_{0,LED} + N_{0,true} \times P_{>threshold}
\end{equation}

\noindent 
where $N_{0,true} \times P_{>threshold}$ is the number of events without any LED photons detected that still have an amplitude above threshold. Plugging equations \ref{eq:pthres} and \ref{eq:n0true} together and isolating $N_{0,true}$, we get: 

\begin{equation}
N_{0,true} = N_{0,LED} \times \frac{N_{tot,ped}}{N_{0,ped}}
\end{equation}

\noindent 
$\frac{N_{tot,ped}}{N_{0,ped}}$ is therefore the correction factor that allows us to properly correct for dark noise pulses which is particularly important for SiPM devices. From Poisson statistics, one then obtains:

\begin{equation}
<N_{ph,detected}> = - Log_{e}\left(\frac{N_{0,true}}{N_{tot,LED}}\right)
\end{equation}

\noindent 
And we finally derive the photon detection efficiency with:

\begin{equation}
PDE = \frac{<N_{ph,detected}>}{<N_{ph,incoming}>}
\end{equation}

\noindent 
We note that because this method relies on the counting of events which are in the pedestal peak (zero PE), a low photon flux onto the photosensor is required so that sufficient events are detected as part of the zero PE peak. It also naturally bypasses the issue of crosstalk and afterpulsing as they do not affect pedestal events whatsoever.

In our setup, we use 5 different combinations of LED with wavelength filters to achieve narrow emission spectra at 375, 400, 455, 500 and 590~nm ($\sim \pm10-15$~nm).

\subsubsection{Results}
\label{sec:pde_results}

\begin{figure}[t]
\centering
\includegraphics[height=2.05in]{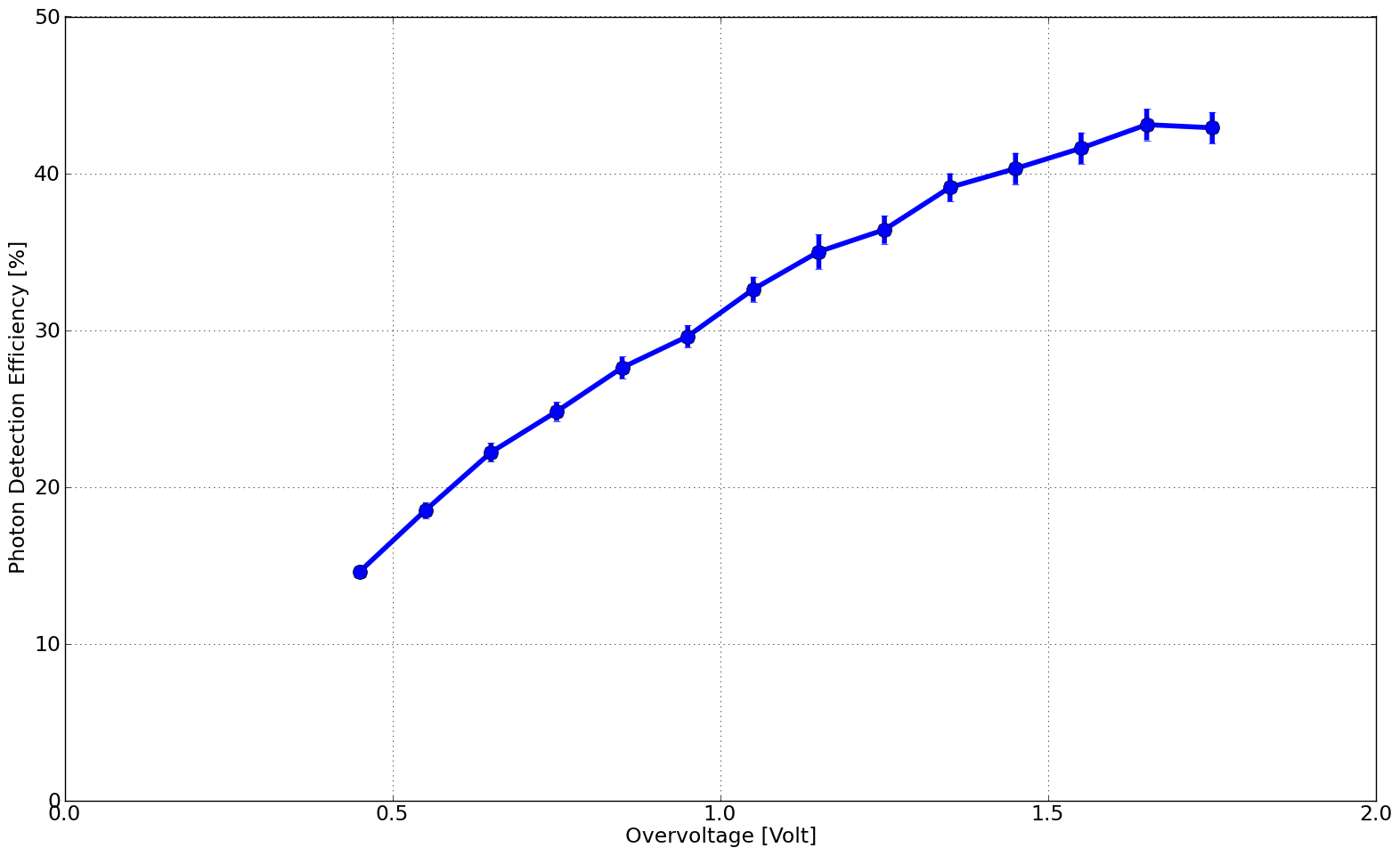}
\includegraphics[height=2.05in]{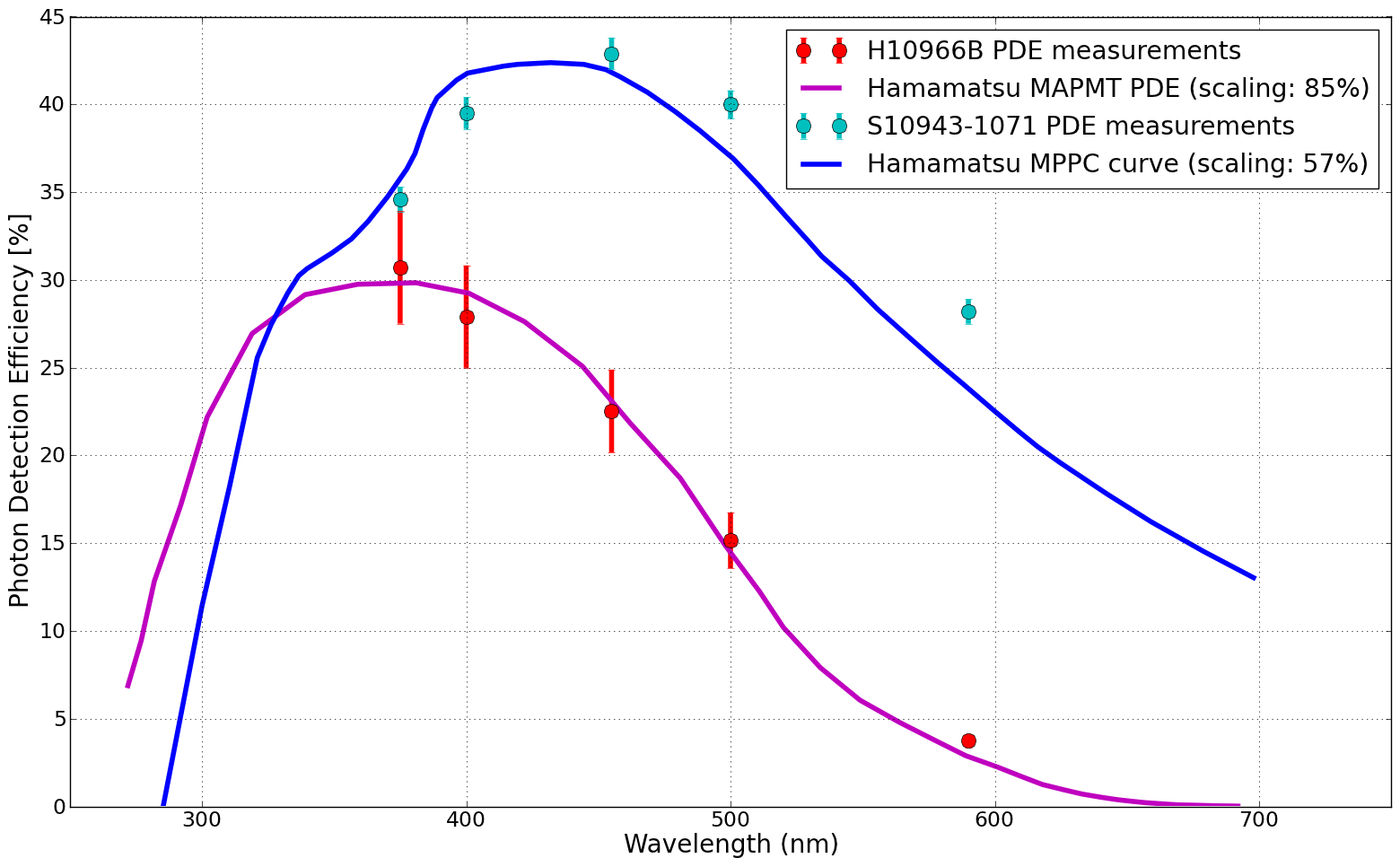}
\caption{{\bf Left}: PDE as a function of overvoltage for MPPC S10943-1071 (455~nm and -5$^{\circ}$C). Although only a single temperature is shown, we find that PDE is temperature independent as long as parametrized by overvoltage. Error bars only represent the statistical uncertainty of our measurements. {\bf Right}: Comparison of MAPMT H10966B (at 1200~V) and MPPC S10943-1071 (at 1.65~V overvoltage) PDE as a function of wavelength. Error bars on MPPC data points represent the statistical uncertainty of our measurements while MAPMT error bars represent the standard deviation of the PDE measurements obtained on all 64 pixels in the tube (statistical uncertainty of each measurement is much smaller). Solid curves give an approximate fit to the data using a scaled version of the efficiency curves provided by Hamamatsu for these devices. The 85\% scaling is an approximate estimate of the collection efficiency (which is not included in the Hamamatsu quantum efficiency curve) although it comes with unknown errors due to systematic uncertainties that exist between the two measurements. MPPC scaling by 57\% is due to Hamamatsu PDE measurement being overestimated due to over-counting from crosstalk and afterpulses (which our method does not suffer from).}
\label{pde_sipm_mapmt}
\end{figure}

SiPM photon detection efficiency is strongly dependent on the applied overvoltage (Figure \ref{pde_sipm_mapmt}, left panel). Higher overvoltages lead to higher electric fields inside the depleted region which increase the chances for an electron-hole pair to initiate an avalanche (higher $P_{avalanche}$). Because PDE is the most critical parameter for IACTs, we intend to operate SiPM devices at the highest overvoltage possible as long as noise (dark rate, cross-talk, afterpulsing) can be kept at an acceptable level (see Section \ref{sec:noise}).

At very high overvoltages, our measurement technique fails because the high noise level prevents us from cleanly resolving the 1~PE peak above pedestal. This is for example the reason that no data points are shown beyond overvoltage $\sim 1.75$~V for MPPC S10943-1071 (Figure \ref{pde_sipm_mapmt}, left panel). We note though that this experimental limitation occurs when the noise is at a level that would not be acceptable for our SC-MST application. 
We expect the PDE to plateau at high overvoltage when $P_{avalanche}$ approaches 100\%. Although this plateau cannot be clearly observed for MPPC S10943-1071, we do observe it on more recent low noise SiPMs (with low crosstalk and afterpulsing) indicating that Geiger avalanches are being generated with full efficiency from an initial electron-hole pair.

Figure \ref{pde_sipm_mapmt} (right panel) compares the PDE of MAPMT H10966B and MPPC S10943-1071. Interestingly, we find higher efficiencies for MPPC even at the lowest wavelength measured (375~nm). Using scaled versions of Hamamatsu efficiency curves (see caption of Figure \ref{pde_sipm_mapmt} for details), we estimate MPPC S10943-1071 has $\sim 60\%$ higher efficiency than MAPMT after convolution on the Cherenkov spectrum (Figure \ref{cherenkov_nsb}). The largest PDE improvement however occurs at high wavelength where the NSB flux is the strongest. Due to this higher red-infrared efficiency, NSB rate is expected to be $\sim 3.5$ higher for MPPC S10943-1071 than MAPMT H10966B. This is an important issue for us as it directly impacts how low of an energy threshold can be achieved with our SC telescope.

Because of its dual mirror design, the Schwarzschild-Couder optics focuses photons onto the focal plane with high incidence angles (up to $\sim 65^{\circ}$ from normal incidence) which makes the use of interference optical filters (to block red-IR photons) on the camera entrance window challenging. Instead, we are investigating the use of a special coating (dielectric550) on the primary mirror which would essentially filter out light above 550~nm. This would reduce the SiPM NSB rate almost by a factor of $\sim 2$ while only reducing Cherenkov efficiency by $\sim 15\%$. Another alternative would be to reduce the red-IR sensitivity of SiPMs while retaining (or improving) their blue-UV sensitivity.

\subsection{Off-axis Efficiency}
\label{sec:angular_response}

\begin{figure}
\floatbox[{\capbeside\thisfloatsetup{capbesideposition={left,center},capbesidewidth=8cm}}]{figure}[\FBwidth]
{\caption{Angular efficiency of Hamamatsu MAPMT H10966B and Excelitas SiPM C30742-33-050-C at 420~nm. The relative transmittance curves are shown for comparison: air-glass-photocathode ($n_{glass} \sim 1.5$, $n_{photocathode} \sim 2.7$) and air-epoxy-silicon ($n_{epoxy} \sim 1.55$, $n_{Si} \sim 3.4$). SiPM results are consistent with off-axis light losses coming purely from light reflections. However MAPMT shows significant additional losses above $\sim 40^{\circ}$.}\label{angular_response}}
{\includegraphics[width=9cm]{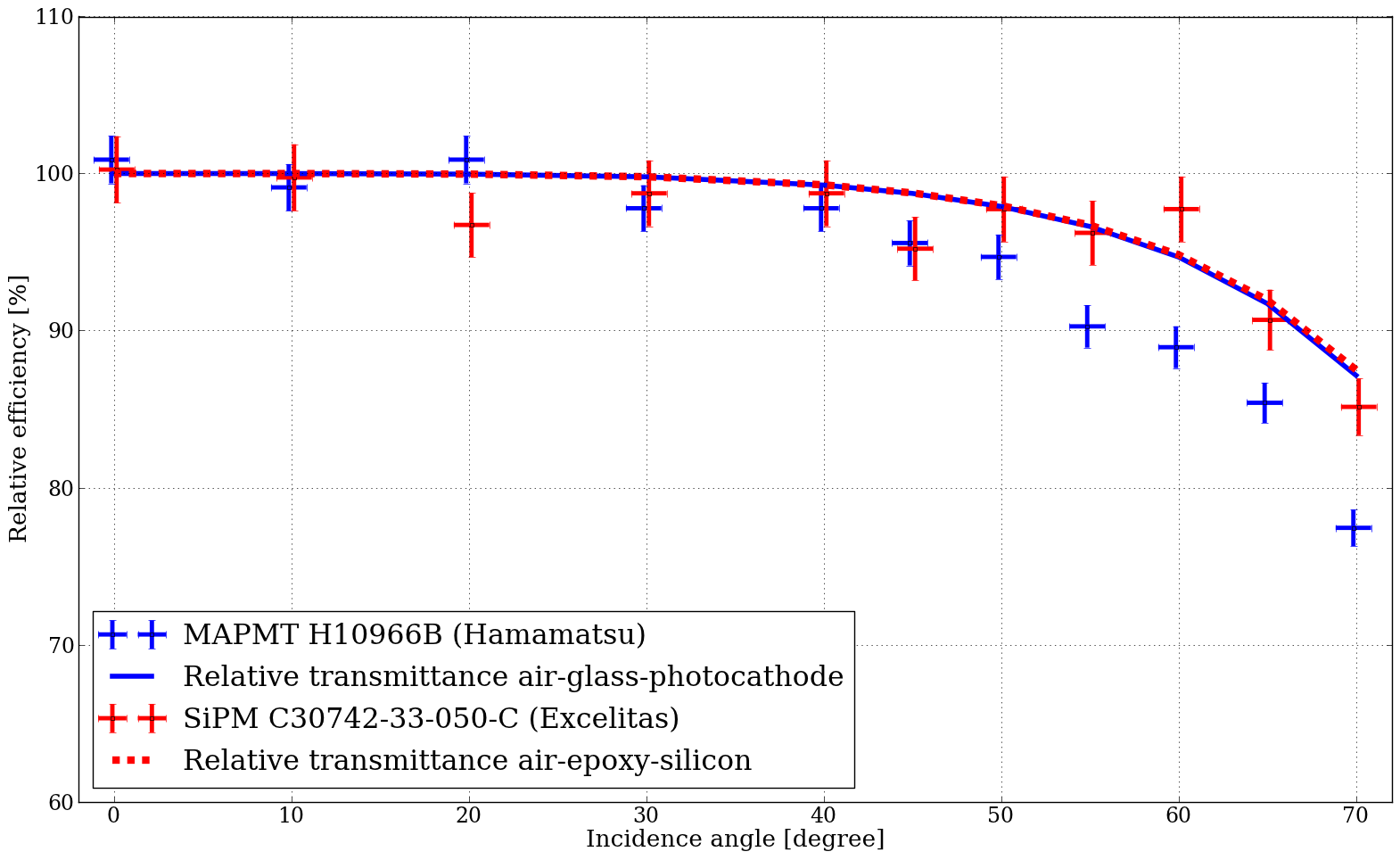}}
\end{figure}

As mentioned in the previous section, the SC optical design has a large fraction of photons which arrive onto the focal plane with large incidence angles. It is therefore critical to evaluate the off-axis efficiency of our photosensors. We measured the relative efficiency of both Excelitas C30742-33-050-C  and MAPMT H10966B up to angles of $70^{\circ}$ at 420~nm (Figure \ref{angular_response}). We find that the efficiency loss at large angles are compatible with the reflection losses from the entrance optical interface for SiPM but that significant additional losses are present for MAPMT above $\sim 40^{\circ}$. We are currently studying the wavelength dependence of this effect.
When convolving the relative efficiency measured with the photon angular distribution in the SC-MST focal plane, the result is an overall efficiency loss of $\sim 2.5\%$ and $\sim 5\%$ for SiPM and MAPMT respectively.

\subsection{Dead area}
\label{sec:dead_area}

The photosensor effective area only covers a fraction of the focal plane surface.
Compact packaging of pixels into large pixel array modules is needed. The MAPMT module already achieves this although we find that inefficiencies at the pixel boundaries result in $\sim 5\%$ efficiency loss.
A compact arrangement of pixel array modules onto the focal plane also needs to be achieved. $\sim 10\%$ efficiency loss occurs by stacking MAPMT modules due to $\sim 1.5$~mm dead band at the edge (mechanical tolerances should only contribute a small fraction).
SiPM pixel signals are currently routed to the backside PCB via wire bonding technology taking some real estate which is essentially lost for the purpose of light detection. Hamamatsu is currently proposing Through Silicon Via technology which has the potential of completely removing the need for this dead area. Excelitas is likewise working on a dense packaging approach.
We are currently working with manufacturers to limit these dead areas as much as practically feasible.

\section{Sources of Noise}
\label{sec:noise}

There are multiple sources of noise that are intrinsic to MAPMT and SiPM devices. In this section, we will only address the most important ones to consider for the Cherenkov telescope application.

\subsection{Dark Noise}
\label{sec:dark_noise}

\begin{figure}[t]
\centering
\includegraphics[height=2in]{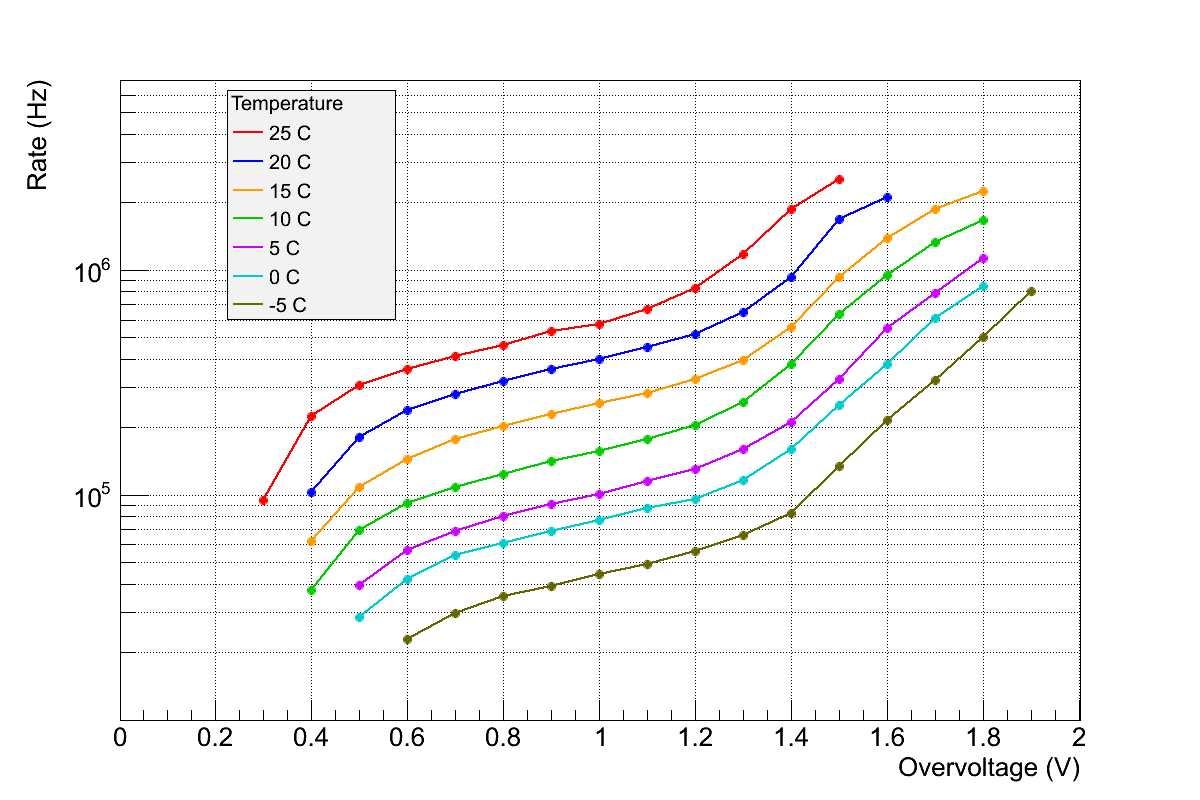}
\includegraphics[height=2in]{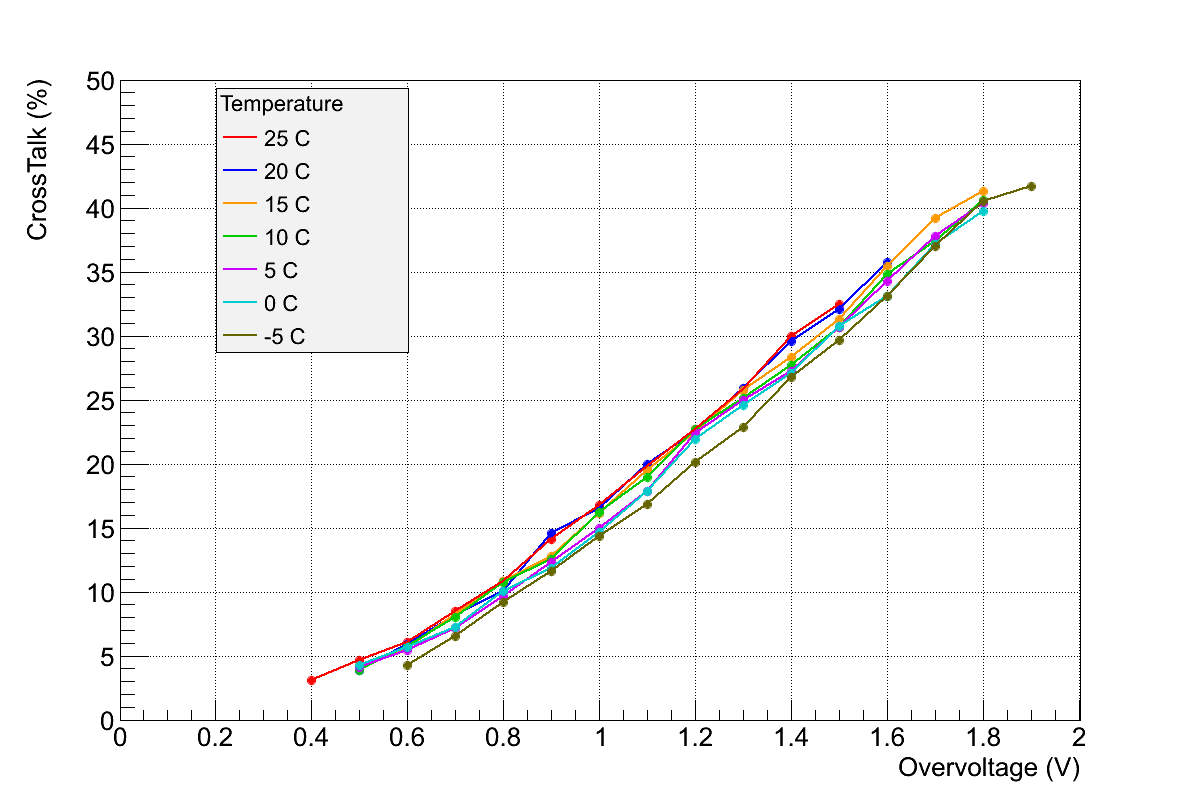}
\caption{Measurements for MPPC S10943-1071. {\bf Left}: Dark rate as a function of temperature and overvoltage. The drop in rate at very low overvoltage is due to the gain being too low for the 1~PE amplitude to remain above the discriminator threshold used to trigger pulses. The pixel size is 3x3~mm$^2$, so the dark rate is $\sim 100$~kHz/mm$^2$ at 25~$^{\circ}$C and recommended operating overvoltage (1.2~V). This can be reduced by a factor of $\sim 2-3$ for every 10~$^{\circ}$C decrease in temperature. {\bf Right}: Cross-talk probability as a function of overvoltage is essentially temperature independent. We find cross-talk probability of $\sim 22\%$ and $\sim 36\%$ at operating overvoltage (1.2~V) and highest PDE overvoltage (1.65~V) respectively for this device.}
\label{rate_xtalk}
\end{figure}

Thermal excitation inside the MAPMT photocathode or SiPM depleted region can bring an electron to the conduction band leading to a so-called dark pulse which is indistinguishable from a photoelectron pulse. Dark pulses are therefore an irreducible source of noise in our detector. This dark noise (a.k.a. thermal noise) is a great source of concern for applications that need to operate in an extremely low noise environment. However, IACTs operate in a naturally noisy environment because of the surrounding night sky background. Indeed, even though sky darkness is one of the prime criteria for IACT site selection, the pollution from NSB photons remains the main limiting factor in determining the instrument low energy threshold. All sources of random noise (such as dark noise) should therefore be at a level significantly below the expected NSB rate not to affect our telescope performance.

From simulation of the SC-MST telescope performance with the NSB spectrum of a site in Namibia, one of the darkest CTA sites being considered, we expect a rate of $\sim 5-20$~MHz/pixel for MAPMT H10966B and $\sim 20-80$~MHz/pixel for MPPC S10943-1071 purely from NSB photons ($\sim 10-40$~MHz if we were to use some sort of filter to cut off $>550$~nm wavelengths) with the pixel size being $\sim$ 6x6 mm$^2$. The factor of $\sim 4$ variation in the NSB rate results from different brightness in  different parts of the sky (galactic fields being typically significantly brighter than extragalactic fields).

Thanks to its high work function (requiring large thermal fluctuations to emit a photoelectron into the vacuum), MAPMT enjoys a low level of dark rate, on the order of few tens of Hz per pixel at 25 $^{\circ}$C for H10966B. On the other hand, SiPMs have typical dark rate of $\sim 100-200$~kHz/mm$^2$ at operating voltage and 25~$^{\circ}$C which means that for a $6\times6$~mm$^2$ pixel, the dark rate is expected to be $\sim 3-6$~MHz. Although smaller than the expected NSB rate, this is not a completely negligible component, especially given the fact that we will operate the SiPM slightly above operating voltage $V_{op}$ in order to achieve the highest PDE. By lowering the temperature by 10$^{\circ}$C, a reduction in the dark rate by a factor $\sim 2-3$ can be achieved (Figure \ref{rate_xtalk}, left panel). Since we plan to regulate the temperature of the SC-MST camera in any case, we will choose an operating temperature which yields reasonable dark rate, probably in the range of $\sim 5-15^{\circ}$C.

\subsection{SiPM cross-talk}
\label{sec:xtalk_sipm}

\begin{figure}[t]
\centering
\includegraphics[height=2.3in]{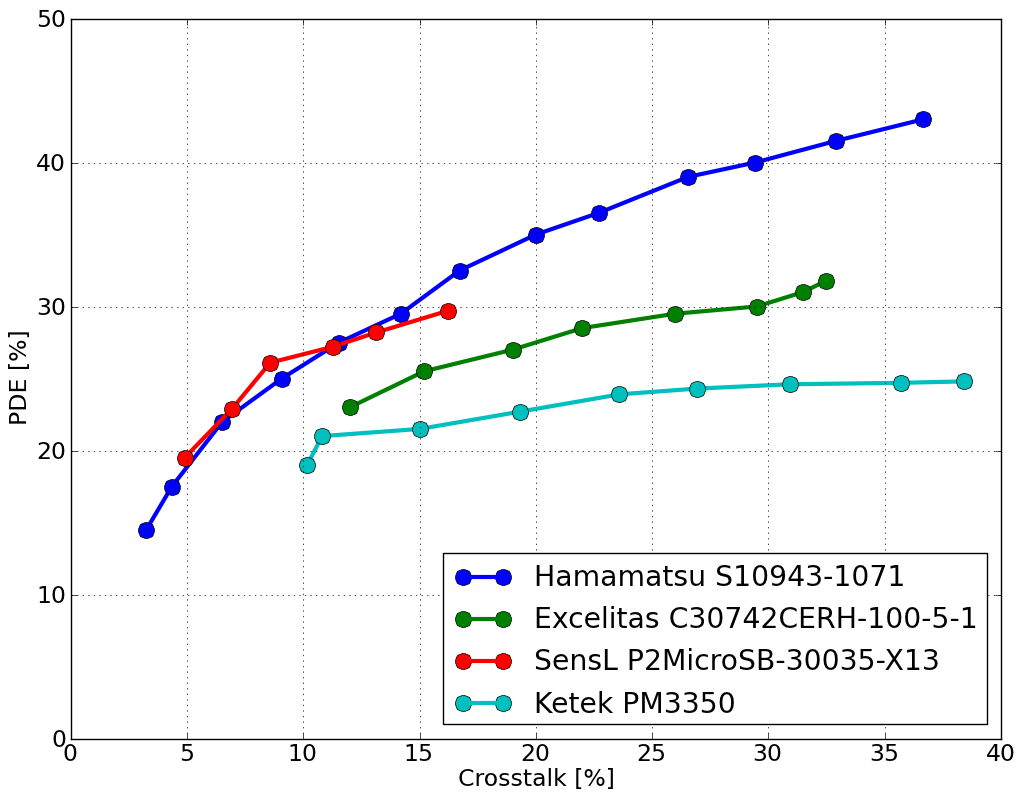}
\includegraphics[height=2.3in]{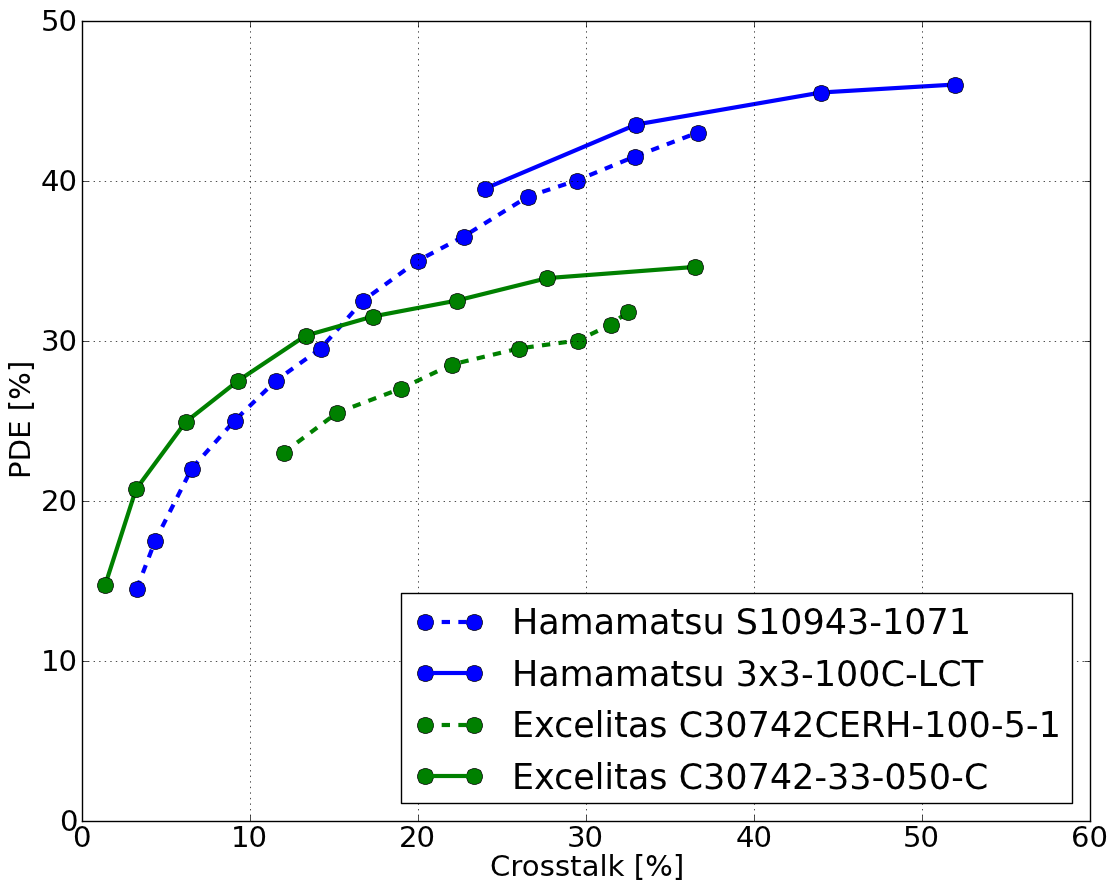}
\caption{Results in the PDE (at 455~nm) and cross-talk phase space. An optimum device for the IACT application would reside in the upper left region of these plots. {\bf Left}:  Results from the best device from each manufacturer before the release of low cross-talk devices. {\bf Right}: New Hamamatsu (MPPC 3x3-100C-LCT) and Excelitas (C30742-33-050-C) devices achieve lower cross-talk for fixed PDE. The improvement is only marginal for the Hamamatsu device but Excelitas achieves a factor of $\sim 2-3$ reduction in cross-talk for a fixed PDE performance. Although not yet at the desired level for CTA, these are definite steps in the right direction.}
\label{xtalk_pde}
\end{figure}

Cross-talk refers to very different phenomena in SiPM and MAPMT. This section addresses cross-talk in SiPM devices which after PDE is the most critical characteristics to consider for IACT application.

A by-product of a Geiger avalanche discharge is infrared photons which can propagate outside the cell and trigger surrounding cells via various mechanisms \cite{Buzhan09}. The probability for this to happen is referred to as the cross-talk probability $P_{crosstalk}$. Light propagation inside the silicon substrate being essentially instantaneous, the output pulse will be the superposition of the signals from the original photo-triggered cell and the cross-talk cells. Cross-talk is in particular the reason behind multiple PE amplitude pulses in Figure \ref{pulse_shape} and \ref{pulse_height} (right panel).

This effect is of paramount importance for Cherenkov telescopes as it gives a mechanism for random NSB or dark pulses to acquire arbitrarily large amplitudes, forcing us to set a high pixel threshold to keep the accidental trigger rate at a reasonable level for the data acquisition system. And as we have now mentioned multiple times, an increase in pixel threshold has a direct impact on the low energy threshold of the array. Monte-Carlo simulations of the SC-MST design reveals that everything being the same (in particular PDE), a cross-talk probability $< 5\%$ would need to be achieved in order to keep the low energy threshold within $\sim 20 \%$ of its optimal value ({\it i.e.} when $P_{crosstalk} = 0$).

The cross-talk probability can easily be calculated from the SiPM pulse height distribution (Figure \ref{pulse_height}, right panel) as:

\begin{equation}
P_{crosstalk} = \frac{N_{>1.5pe}}{N_{>0.5pe}}
\end{equation}

The cross-talk probability correlates with overvoltage (Figure \ref{rate_xtalk} ) because it scales with the number of photons produced inside an avalanche (proportional to the gain and thus the overvoltage) as well as the probability of one of these photons to trigger a neighboring cell (proportional to PDE and thus the overvoltage). 

Figure \ref{xtalk_pde} summarizes the performance of SiPM devices we have obtained from different manufacturers in the phase space of the two most important parameters for Cherenkov telescopes: PDE and cross-talk. Our wish for CTA is to find a high PDE and low cross-talk SiPM device ({\it i.e.} living in the upper left part of the figure).

Hamamatsu and Excelitas are recently producing devices with cross-talk suppression using trenches, {\it i.e.} etched grooves around each cell coated with a reflecting material for optical isolation from neighboring cells. Although the trenches decrease somewhat the fill factor of the SiPM pixel, they allow operation at higher overvoltage thanks to a reduced noise level (also helped by reduced afterpulsing, see section \ref{sec:afterpulsing}). An interesting consequence is that for some of these devices we see a clear PDE versus overvoltage plateau indicating full efficiency at generating a Geiger avalanche from an initial electron-hole pair ($P_{avalanche} \sim 100\%$).
The right panel of Figure \ref{xtalk_pde} shows the improvement we have observed with the first generation of such devices in the PDE versus cross-talk phase space which is key for the CTA application. Excelitas in particular was able to achieved a cross-talk reduction of a factor of $\sim 2-3$ at fixed PDE performance which is quite encouraging. 

For a given device, there is also the question of which operating overvoltage will yield the lowest SC-MST energy threshold (a compromise between low cross-talk and high PDE). Preliminary Monte-Carlo simulations indicate that high PDE is the prime criterion even if it comes at the cost of high cross-talk probability. We emphasize however that for fixed PDE, $P_{crosstalk} < 5\%$ is highly desirable.

\subsection{MAPMT crosstalk}
\label{sec:xtalk_mapmt}

There are mainly two different mechanisms by which a MAPMT signal (or partial signal) can be detected by other pixels than the one where the photon was converted to a photoelectron. 

If the photoelectron exits the photocathode with large transverse momentum, there is a chance the electric field lines will not be able to focus it on the first dynode. It will instead drift to a neighboring pixel where it will be amplified and eventually detected. This type of cross-talk is usually referred to as optical cross-talk although it is not purely an optical effect. It is an issue for our application as it introduces an uncertainty on the photon impact point on the focal plane. As expected, the intensity of this effect depends on the exact location where the photon hits the photocathode. We find it is at the few percent level at the center of a pixel but can reach up to 50\% at the very edge of a pixel. We estimate that the combined effect is an overall increase in photon position uncertainty by $\sim 15\%$.

Another type of cross-talk we identified is most clearly visible when a large amplitude signal illuminates a small part of the module (as will be the case with EAS). For example, if a single pixel is illuminated with a large amplitude pulse, we observe a `sympathetic' signal of amplitude $\sim 0.5\%$ of the main signal in all the other pixels of the tube independent of their location with respect to the pixel being illuminated. We suspect this is due to electronic cross-talk through the common high voltage line although we have not investigated this effect in full detail.

\subsection{Afterpulsing}
\label{sec:afterpulsing}

Afterpulsing refers to very different processes inside PMT or SiPM devices.

In PMT technology, positive ions can be created through ionization by avalanche electrons of residual gas inside the vacuum. The ion will then drift back toward the photocathode where it generates multiple secondary electrons which result in a delayed large amplitude pulse. In MAPMT H10966B, we find afterpulsing probabilities lower than 0.1\% which is deemed acceptable for the Cherenkov telescope application.

With SiPM, charge carriers can be trapped by impurities with intermediate energy levels during the avalanche process. When released, they will generate delayed avalanches which will contribute to the total noise observed. Recent devices from Hamamatsu (ex: MPPC 3x3-100C-LCT) have significantly reduced afterpulsing rate thanks to improved purity of the silicon substrate which allows operation at higher overvoltages.




\section{Aging}
\label{sec:aging}

A major drawback of PMT technology is its performance degradation as it is exposed to large amounts of light. The weak point in the tube is the last dynode which undergoes heavy bombardment from electron avalanches. This is particularly detrimental to the Cherenkov application as it makes operation during bright moonlight period very difficult if not impossible. Current IACT instruments have developed work-arounds by operating PMTs under low gain configuration (by reducing the high voltage) or installing ultraviolet filters ({\it i.e.} filtering out high wavelength typically above $\sim 400$~nm) in front of PMTs under moonlight. However these come at the price of somewhat degrading the telescope performance.

We carried out aging tests on the MAPMT H10966B to determine how critical performance parameters evolve as a function of total charge deposited on the anode. For this, we uniformly illuminated pixel 15 extending the illumination to the neighbor pixels to properly include the optical cross-talk effect. We continued aging until the total charge deposited on pixel 15 was $\sim 70$~Coulombs which would correspond to $\sim 35$~years of SC-MST operation ($\sim 2$~Coulombs/season). We regularly interrupted the aging process to measure the gain, afterpulsing and PDE of pixel 15 as well as two reference pixels that were not being aged (protected by a mask). Even after 70~Coulombs of deposited charge, we found no sign of degradation of any of the parameters being monitored. We point out however, that because of the poorly resolved single PE peak (see section \ref{sec:gain}), the systematic uncertainty on the gain measurement was found to be at the level of $\sim \pm 20\%$ (estimated from the reference pixels). In any case, this is an indication that aging might not be as much of a problem in MAPMT as we originally thought.

Although we have not yet carried out aging tests on any SiPM device, the expectation is that silicon-based technology should be very robust against high levels of light exposure. We will soon carry out tests to verify this hypothesis.


\section{Conclusion}
\label{sec:conclusion}

In preparation for CTA, we are carefully investigating the option of a using novel Schwarzschild-Couder telescope design for some of the medium size telescopes which will compose the core of the array. We find that both MAPMT and SiPM appear as viable technologies to be used as photodetectors for this optical design since no particularly alarming features was discovered in either of them during the course of our studies.

Although no show-stopper appears to be in the way, the MAPMT H10966B has known limitations coming mostly from PMT technology: fragility, high voltage operation, aging and limited PDE. SiPMs on the other hand do not suffer from any of these although they have the drawback of requiring precise temperature regulation of the camera. PDE, the most important parameter for a Cherenkov telescope, is in particular found to be significantly better for the most efficient SiPM devices than for MAPMT H10966B, despite the fact that a large margin for improvement still exists. Cross-talk is currently the main source of worry with SiPMs although significant improvements have recently been made by some manufacturers and further technological advances are expected to continue reducing this noise source.

We are currently designing a SC-MST prototype to be built at the Fred Lawrence Whipple Observatory in southern Arizona with first light expected in fall 2015. For this prototype, SiPM technology has been selected to populate the camera at the focal plane. The prototype will serve as a proof-of-concept for the Schwarzschild-Couder design. It will also give us the necessary experience to operate SiPM devices in the field. We continue in parallel to collaborate with SiPM manufacturers in order to further improve SiPM technology with the goal of having a close to ideal photodetector when CTA comes online.

\acknowledgments     

This work would not have been possible without the generosity of SiPM and MAPMT manufacturers  ({\it Excelitas}, {\it Hamamatsu Photonics}, {\it SensL}, {\it FBK AdvanSiD}, {\it Ketek}) that have provided us with device samples to carry out our investigations. The authors gratefully acknowledge support from the following agencies and organizations:
National Science Foundation; University of California; Institute for Nuclear and Particle Astrophysics (INPAC-MRPI program); Georgia Institute of Technology; Science and Technologies Facilities Council, UK. This work was also supported by JSPS KAKENHI Grant Number 23244051.


\bibliography{report}   
\bibliographystyle{spiebib}   

\end{document}